\documentclass[a4paper,onecolumn,10pt]{quantumarticle}
\pdfoutput=1
\usepackage{amsthm}
\usepackage[utf8]{inputenc}
\usepackage[english]{babel}
\usepackage[T1]{fontenc}
\usepackage{amsmath}
\usepackage{hyperref}
\usepackage[numbers]{natbib}

\usepackage{tikz}
\usepackage{lipsum}
\usepackage{mathrsfs}
\newcommand{\e}{{\mathrm e}}
\newcommand{\Ccal}{{\mathcal C}}

\newcommand{\eq}[1]{\begin{equation} #1 \end{equation}}
\newcommand{\tr}{\mathrm{Tr}}

\newcommand{\eqarray}[1]{\begin{eqnarray} #1 \end{eqnarray}}
\newcommand{\ket}[1]{\vert #1 \rangle}
\newcommand{\bra} [1] {\langle #1 \vert}

\newcommand{\ketbra}[2]{| #1 \rangle \langle #2 |}

\newcommand{\mean}[1]{\langle #1 \rangle}

\usepackage{graphicx}% Include figure files
\usepackage{dcolumn}% Align table columns on decimal point
\usepackage{bm}% bold math
\usepackage{color}
\usepackage{soul}
\usepackage[dvipsnames]{xcolor}
\usepackage{mathtools}
\usepackage{dsfont}
\usepackage{amssymb}
\usepackage{tcolorbox}
\usepackage{makecell}
\usepackage{tocloft}
\usepackage{mdframed}
\usepackage{makecell}
\usepackage[table]{xcolor}

\usepackage{pifont}
\usepackage{amssymb}
\usepackage{titlesec}
\usepackage[margin=1in]{geometry} % Set the margin width
\newtheorem{theorem}{Theorem}[section]

\newtheorem{lemma}[theorem]{Lemma}

\begin{document}

\title{Quantum Wasserstein distance for Gaussian states}
\author{Anaelle Hertz}
\affiliation{National Research Council of Canada, 100 Sussex Drive, Ottawa, Ontario K1N 5A2, Canada}
\email{anaelle.hertz@nrc-cnrc.gc.ca}
\orcid{0000-0002-9608-512X}
\author{Mohammad Ahmadpoor}
\affil{Department of Mathematics and Statistics, University of Ottawa, 150 Louis-Pasteur Pvt, Ottawa, Ontario, K1N 6N5 Canada}
\orcid{0000-0001-6902-1916}
\author{Oleksandr Dzhenzherov}
\affiliation{Department of Physics,  University of Ottawa, 25 Templeton Street, Ottawa, Ontario, K1N 6N5 Canada}
\author{Augusto Gerolin}
\affil{Department of Mathematics and Statistics and Department of Chemistry and Biomolecular Sciences,  University of Ottawa, 150 Louis-Pasteur Pvt, Ottawa, Ontario, K1N 6N5 Canada}
\affiliation{Nexus for Quantum Technologies, University of Ottawa, 25 Templeton Street, Ontario, K1N 6N5 Canada}
\orcid{0000-0002-2573-900X}
\affiliation{Instituto de Matem\'atica Pura e Aplicada, 
Dona Castorina Road, 110,
Jardim Botânico, Rio de Janeiro, RJ22460-320,
Brazil}
\author{Khabat Heshami}
\affiliation{National Research Council of Canada, 100 Sussex Drive, Ottawa, Ontario K1N 5A2, Canada}
\affiliation{Department of Physics,  University of Ottawa, 25 Templeton Street, Ottawa, Ontario, K1N 6N5 Canada}
\affiliation{Institute for Quantum Science and Technology, Department of Physics and Astronomy, University of Calgary, Alberta T2N 1N4, Canada}
\orcid{0000-0003-3864-1930}
\maketitle
\begin{abstract}
Optimal transport between classical probability distributions has been proven useful in areas such as machine learning and random combinatorial optimization. Quantum optimal transport, and the quantum Wasserstein distance as the minimal cost associated with transforming one quantum state to another, is expected to have implications in quantum state discrimination and quantum metrology. In this work, following the formalism introduced in [De Palma, G. and Trevisan, D. Ann. Henri Poincaré, {\bf 22} (2021), 3199-3234] to compute the optimal transport plan between two quantum states, we give a general formula for the Wasserstein distance of order 2 between any two one-mode Gaussian states. We discuss how the Wasserstein distance between classical Gaussian distributions and the quantum Wasserstein distance by De Palma and Trevisan for thermal states can be recovered from our general formula for Gaussian states. This opens the path to directly compare various known distance measures with the Wasserstein distance through their closed-form solutions.
\end{abstract}

\section{Introduction}
Optimal transport theory provides a framework for determining the optimal way to transform one probability distribution into another. Originating in 1781 with Monge's formulation \cite{Monge1781} to find the minimal cost to move a mass distribution in one configuration into another, the theory was firmly established by Kantorovich in 1942 \cite{Kantorovich1942} by introducing the notion of coupling through a joint distribution that links the source and target measures. Instead of assigning each point $x$ to a single destination $y$, Kantorovich considered all possible ways of pairing mass from the source with mass in the target, encoded by a probability measure on the product space. This transformed the problem into a linear minimization over a convex set. The result of minimizing the cost over all couplings defines the so-called Wasserstein distance. The optimal transport theory spurred the development of several novel tools with applications in machine learning \cite{PeyreCuturi2019COT}, random combinatorial optimization \cite{Caracciolo2014Scaling}, and reformulating partial differential equations \cite{Otto2001Geometry}.

Classical transport plans are interpreted as the physical operations required to redistribute resources (mass, fluid, probabilities, etc.) from a source to a target and the Wasserstein distance as the minimal cost of such operations \cite{Villani2009}. It is natural to seek formulations of optimal transport in quantum systems and explore their foundation and any potential applications. Several approaches have been proposed to generalize optimal transport to the quantum setting. Zyczkowski and S{\l}omczy{'n}ski \cite{ZyczkowskiSlomczynski1998} computed the Wasserstein distance between Husimi probability distributions, providing an early phase-space-based approach. In the context of free probability, Biane and Voiculescu \cite{BianeVoiculescu2001} introduced an analogue of the Wasserstein metric. Carlen, Maas, Datta, and Rouzé \cite{CarlenMaas2014,CarlenMaas2017,CarlenMaas2020, RouzeDatta2019,DattaRouze2020,ChenGeorgiouTannenbaum2018a,ChenGeorgiouTannenbaum2018b} defined a notion of distance describing the evolution of a density matrix under Lindblad-type dynamics. Golse, Mouhot, Paul, and Caglioti \cite{GolseMouhotPaul2016,CagliotiGolsePaul2018preprint,Golse2018,GolsePaul2017,GolsePaul2018,CagliotiGolsePaul2020} have a static interpretation and defined the quantum analogue of the Kantorovitch problem on density operators, employing phase-space techniques. De Palma and Trevisan \cite{DePalma2021} formulated a quantum Wasserstein distance using quantum channels, while De Palma, Marvian, Trevisan, and Lloyd \cite{DePalmaMarvianTrevisanLloyd2021} generalized the Wasserstein distance of order 1 to n-qudit systems. Finally, Bistron, Cole, Eckstein, Friedland, and Zyczkowski \cite{FriedlandEcksteinColeZyczkowski2022,ColeEcksteinFriedlandZyczkowski2021,BistronEcksteinZyczkowski2023} proposed a quantum Wasserstein distance based on an antisymmetric cost function. Detailed reviews of all different approaches to the quantum optimal transport can be found 
in \cite{MaasRademacherTitkosVirosztek2024,beatty2025wasserstein}.

In classical optimal transport, every transport plan corresponds uniquely to a stochastic map, giving the Wasserstein distance a direct operational meaning: it quantifies the minimal “effort” required to implement a physical transformation of the system. The formulation introduced by De Palma and Trevisan \cite{DePalma2021,DePalma2024} is of particular interest as quantum channels, completely positive, trace-preserving maps acting on the space of density operators, are the quantum analogue of the transport plan. This feature distinguishes their approach from other quantum transport proposals and allows the distance to retain a clear operational interpretation in terms of physically implementable transformations on the quantum system and the distance associated with such transformations. While their framework is mathematically well established, it has so far only been applied to compute the Wasserstein distance between thermal states. In this paper, we extend their approach by explicitly computing the distance between arbitrary one-mode Gaussian states, thereby providing a broader class of examples and making the operational interpretation applicable to a wider set of quantum states. 
The main result of the paper is the following:
\begin{mdframed}[backgroundcolor=quantumviolet!5!,linecolor=quantumviolet!5!]
\begin{theorem}
\label{thm:Mainresult}
    Let $A$ and $B$ be the covariance matrices of two one-mode Gaussian states $\rho_A$ and $\rho_B$, then the Wasserstein distance  is given by
  $$
D^2(\rho_A,\rho_B)\!=\! \frac12\tr[A+B]-\frac12\sqrt{\frac{4\sqrt{\det A}\sqrt{\det B}-2|\sqrt{\det A}\!-\!\sqrt{\det B}|-1
 }{\sqrt{\det A}\sqrt{\det B}}}\tr[\sqrt{\!\sqrt{B}A\sqrt{B}}]
$$
\end{theorem}
\end{mdframed}

The paper is structured as follows. Section~\ref{sec:optimaltransport} introduces the definition of the quantum Wasserstein distance based on quantum channels, establishing the operational framework used throughout. Section~\ref{sec:Gaussian} reviews the required preliminaries specific to Gaussian states and Gaussian channels. In Section~\ref{sec:QOTforGaussian}, we formally state the problem we want to solve in the context of Gaussian states and review the known case of thermal states. We then start by investigating the case of Gaussian states related by a unitary transformation before solving the general problem and obtaining the Wasserstein distance between any two one-mode Gaussian states in Section~\ref {sec:Wasserstein Gaussian states}. 
Naturally, since the quantum minimal transport refers to a distance, it is important to understand whether it can serve as a tool for distinguishing quantum states. Several works have explored this question under different formulations of the quantum optimal transport theory \cite{DePalma2021,CarlenMaas2020,DattaRouze2020,CagliotiGolsePaul2020,FriedlandEcksteinColeZyczkowski2022}. 
In Section~\ref{sec:distances} we define the most commonly used notions of distinguishability, which open the door for a comparison with the Wasserstein distance.
Section~\ref{sec:conclusion} concludes with a discussion and outlines directions for future work.

\section{Definition of quantum optimal transport}
\label{sec:optimaltransport}

Adopting the formulation of De Palma and Trevisan \cite{DePalma2021,DePalma2024} we characterize quantum optimal transport as follows.
As in the classical definition, one can define  a coupling $\Pi
_\Phi\in \mathcal{H}\otimes\mathcal{H}^*$ between states  $\rho_A\in\mathcal{H}$ and $\rho_B\in\mathcal{H}$ such that 
\eq{\tr_\mathcal{H}(\Pi_\Phi)=\rho_B^T, \qquad\qquad\tr_{\mathcal{H}^*}(\Pi_\Phi)=\rho_A.}
The set of all such couplings is denoted $\mathscr{C}(\rho_A,\rho_B)$.
We associate a quantum transport plan $\Phi$ from $\rho_B$ to $\rho_A$ and then, the coupling state can be written as 
\eq{\Pi_\Phi=\left(\Phi\otimes\mathds{1}_{\mathcal{H}^*}\right)\ket{\ket{\sqrt{\rho_B}}\!}\bra{\!\bra{\sqrt{\rho_B}}},\label{CouplingAsChannel}}
where $\ket{\ket{\sqrt{\rho_B}}\!} \in \mathcal{H}\otimes\mathcal{H}^*$ is the canonical purification of $\rho_B$. The transport plan $\Phi$ is a quantum channel. A one-to-one correspondence between the set of transport plans and the set of quantum channels has been proven in $\cite{DePalma2021}$. 
Among all the possible couplings, we then search for the coupling $\Pi_\Phi$ which minimizes the average cost $\langle C \rangle_{\Pi_\Phi}=\tr[C\Pi_\Phi]$. Different cost functions exist, whose optimization lead to different order of the Wasserstein distance, but here we focus on the quadratic 
cost (that will give rise to the 2-order Wasserstein distance)
\eq{C=\sum_{i=1}^n(\hat R_i\otimes\mathds{1}_{\mathcal{H}^*}-\mathds{1}_\mathcal{H}\otimes \hat R_i^\top)^2,\label{cost}}
where $\boldsymbol{\hat R}=(\hat X_1,\hat P_1,\hat X_2,\hat P_2,\cdots,\hat X_n,\hat P_n)$ are the quadrature observables of the Hilbert space $\mathcal{H}$ and  satisfy the canonical commutation relations\footnote{Note that we employ units in which $\hbar=1$ throughout this paper.} $[\hat{X}_i,\hat{P}_j]=i\delta_{i,j}$ and $[\hat{X}_i,\hat{X}_j]=[\hat{P}_i,\hat{P}_j]=0$. $\boldsymbol{\hat{R}}^\top$ belongs to the conjugated Hilbert space $\mathcal{H}^*$, and the quadratures satisfy the commutations relations $[\hat{X}_i^\top,\hat{P}_j^\top]=-i\delta_{i,j}$ and $[\hat{X}_i^\top,\hat{X}_j^\top]=[\hat{P}_i^\top,\hat{P}_j^\top]=0$. 
The minimum cost is what we call the (squared) Wasserstein distance\footnote{Throughout the paper,  we will refer to the 2-order Wasserstein distance simply as the \textit{Wasserstein distance}.}
\eq{D(\rho_A,\rho_B)^2=\underset{\Pi_\Phi\in\mathscr{C}(\rho_A,\rho_B)}{\text{min}}\tr [C\Pi_\Phi].\label{eq:WassersteinDef}}
The quantum Wasserstein distance is not a genuine metric as it is non-zero for coinciding quantum states. The distance between a state and itself is connected to the Wigner-Yanase skew information \cite{WignerYanase} defined as
 $I(\rho,\hat X)=-\frac12\tr([\sqrt{\rho},\hat X]^2)$.
 It is used to define a quantifier of nonclassicality\footnote{Nonclassicality here is to be understood in the sense that a state does not have a positive Glauber-Sudarshan quasiprobability distribution, that is it
cannot be written as a convex mixture of coherent states \cite{Titulaer}.} such as (for 1 mode) \eq{\mathcal N(\rho)=\frac12(I(\rho,\hat X)+I(\rho,\hat P))\label{SkewInfoWitness}.} By comparing with Ref.~\cite{DePalma2021}, we observe that the Wasserstein distance of a state with itself is $D(\rho,\rho)^2=2 \mathcal N(\rho)$.
Hence, the Wasserstein distance of a state with itself is a quantifier of nonclassicality: if $D(\rho,\rho)^2>1$, then $\rho$ is nonclassical.
Note that the Wigner-Yanase skew information is only a special version of the infinite families of quantum Fisher information. If we define the Fisher information matrix as 
$J(\rho)_{ab}=\langle\!\langle[R_a,\sqrt{\rho}]\ \|\ [R_b,\sqrt{\rho}]\rangle\!\rangle$, then $D(\rho,\rho)^2=\tr J(\rho).$

If one of the state is pure, say $\rho_A=\ket{\psi}\bra{\psi}$ , the only admissible coupling with reduced states $\rho_A,\ \rho_B$ is the separable state
$\Pi=\ketbra{\psi}{\psi}\otimes\rho_B$, with no correlations between the two systems. Hence, the Wasserstein distance reduces to evaluating the cost on this product state;  no optimization is required.
 In particular, assuming both states are pure and centered on the origin, the cost is equal to the total noise (sum of the variances of each quadrature) \cite{Yadin18,Kwon2} which is also equal to the quadrature coherence scale~\cite{Debievre,Hertz}, $\mathcal
 C^2(\rho)=-\frac1{2n\tr\rho^2}\sum_j\tr[\rho,R_j]^2$,  
 and to the total quantum Fisher information \cite{Yadin18,Kwon2}:
\eq{D(\ket{\psi},\ket{\phi})^2=\mean{\hat n}_\psi+\mean{\hat n}_\phi+1=\frac{n}{2}\,\Ccal^2(\Pi),\label{WassersteinPureStates}}
with $\mean{\hat n}$ being the average photon number and $n$ the number of modes.

\section{Preliminaries for Gaussian states}
\label{sec:Gaussian}

Because the quantum Wasserstein distance remains analytically intractable for general quantum states, it is natural to first focus on Gaussian states. Gaussian states occupy a privileged position in continuous-variable quantum information: they are fully characterized by their first (displacement) and second (covariance matrix) moments, their phase-space representation is a Gaussian probability distribution and it remains Gaussian under all Gaussian operations. This combination of mathematical simplicity and physical relevance makes them the canonical testbed for extending classical concepts to the quantum domain. In particular, many quantum distances and entropic quantities have been derived specifically for Gaussian states where the result can be expressed uniquely in terms of the first and second moments of the states (e.g. \cite{PhysRevA.73.062330,PhysRevA.71.032336,Parthasarathy,Seshadreesan,OZAWA2000158,Coles}). Studying the Wasserstein distance for Gaussian states is therefore an essential first step, both because it yields tractable analytic expressions and because it provides a benchmark for understanding how this distance behaves in more general continuous-variable systems.

 \subsection{Phase space formalism and Gaussian states}

To set the stage, we provide a brief overview of the symplectic formalism used for continuous-variable states in quantum optics and focus on Gaussian states. More details can be found, for example, in \cite{weedbrook, Anaellethesis}.

A continuous-variable system is represented by $n$ modes. To each of them are associated the annihilation and creation operators $\hat a_i$ and $\hat a_i^\dag$ verifying the bosonic commutation relation $[\hat a_i,\hat a_i^\dag]=~1$. The elements of the vector of quadratures $\boldsymbol{\hat R}$ (as defined below Eq.~\eqref{cost}), 
can be written as
$\hat{X}_j=\frac1{\sqrt{2}}(\hat a_j+\hat a_j^\dag)$, $ \hat{P}_j=-\frac{i}{\sqrt{2}}(\hat a_j-\hat a_j^\dag)$,  for $ j=1,\cdots,n$.
The first-order moments of a state $\rho$ constitute the displacement vector, defined as 
 $\boldsymbol{d}=\mean{\boldsymbol{\hat R}}=\tr (\boldsymbol{\hat R}\rho)$, 
while the second moments make up the covariance matrix $\gamma$, a real symmetric matrix,  whose elements are given by
\begin{equation}
\gamma_{ij}=\frac12\mean{\{\hat R_i,\hat R_j\}}-\mean{\hat R_i}\mean{\hat R_j}
\label{covmat}	,
\end{equation}
where $\{\cdot,\cdot\}$ represents the anticommutator.  To be the covariance matrix of a valid quantum state, $\gamma$ must satisfy the uncertainty principle 
with
\eq{\gamma\pm\frac{i}{2}\Omega\geq0\qquad\text{with}\qquad
\Omega=\bigoplus_{j=1}^n\begin{pmatrix}0&1\\-1&0	\end{pmatrix}\label{Omega},
}
$\Omega$ being the symplectic form.
According to the Williamson's theorem, every covariance matrix can be brought to a diagonal form $\gamma^\oplus=S\gamma S^\top=\bigoplus_{k=1}^n\nu_k\mathds1_{2\times2}$, with $\nu_k$ being the symplectic values and $S$ a symplectic transformation that preserves the commutation relations, i.e. $S\Omega S^\top=\Omega$ and $\det S=1$. For a one-mode state, the symplectic value is simply equal to $\nu=\sqrt{\det\gamma}$.

Gaussian states are particular as they are fully characterized by their displacement vector $\boldsymbol{d}$ and covariance matrix $\gamma$. The name comes from the fact that their Wigner (and characteristic) functions are a Gaussian function in the phase space.
The most general one-mode Gaussian state can be written as $\rho_G=D(\alpha)U_S\rho_{\rm{th}} U_S^\dag D^\dag(\alpha)$ where $D( \alpha)=\e^{\alpha\cdot a^\dag-\bar{\alpha} \cdot a}$ is the displacement operator and $U_S=R(\theta)S(r)R(\phi)$ with $R(\theta)$ being a rotation and $S(r)=\e^{\frac12(r\hat a^2-r\hat a^{\dag 2})}$ the squeezing operator.
The (one-mode) thermal state $\rho_{\rm{th}}$ has the density matrix
\eqarray{
\rho_{\text{th}} &=& \sum_{n=0}^\infty \frac{\langle\hat{n}\rangle^n}{(\langle\hat{n}\rangle+1)^{n+1}} \, \ket{n} \bra{n}=\sum_{n=0}^\infty  \frac{\left(\nu - \tfrac{1}{2}\right)^n}{\left(\nu + \tfrac{1}{2}\right)^{n+1}}  \, \ket{n} \bra{n},
}
where $\ket{n}$ are the Fock states, $\langle\hat{n}\rangle$ is the mean photon number of the state and $\nu=\langle\hat{n}\rangle+\frac12$ is its symplectic value. Thermal states have the particularity of having a covariance matrix proportional to the identity: $\gamma_{\text{th}}=\left(\bar n+\frac12\right)\mathds{1}=\nu\mathds{1}$.
From this, we can compute the covariance matrix of a centered squeezed thermal state $\rho_{\rm{SqTh}}=S(r)\rho_{\rm{th}} S^\dag(r)$ as
\eq{\gamma_{\text{SqTh}}=
\begin{pmatrix}
    \e^{-r}&0\\0&\e^{r}\end{pmatrix}\gamma_{\text{th}}
    \begin{pmatrix}
    \e^{-r}&0\\0&\e^{r}\end{pmatrix}=\nu   \begin{pmatrix}
    \e^{-2r}&0\\0&\e^{2r}\end{pmatrix}.
\label{gammasqth}}

\subsection{Gaussian channels}
\label{Gaussian channels}

Given an initial Gaussian state of $n$ modes, the evolution due to a deterministic Gaussian CP-map is completely described by two $2n \times 2n$ real matrices $U$ and $V$, which act as follows on the statistical moments \cite{serafini}:
\eq{
\boldsymbol{d} \mapsto U \boldsymbol{d} , 
\qquad
\gamma \mapsto U \gamma U^\top + V.
\label{defchannel}}
The matrices $U$ and $V$ must be such that
\eq{
V \pm \frac{i}{2}\Omega \geq \pm\frac{i}{2} U \Omega U^\top.
\label{ChannelConstraint}}
The output of a Gaussian CP-map (channel) on a subsystem $A$ of a multipartite system in a Gaussian state with first and second moments
\eq{
\boldsymbol{d}_{\mathrm{in}} = \begin{pmatrix} \boldsymbol{d}_A \\ \boldsymbol{d}_B \end{pmatrix}, \qquad 
\gamma_{\mathrm{in}} = \begin{pmatrix} 
\sigma_A & \sigma_{AB} \\
\sigma_{AB}^\top & \sigma_B 
\end{pmatrix} 
}
is the Gaussian state with first and second moments
\eq{
\boldsymbol{d}_{\mathrm{out}} = \begin{pmatrix} U \boldsymbol{d}_A \\ \boldsymbol{d}_B \end{pmatrix}, \ \qquad
\gamma_{\mathrm{out}} = \begin{pmatrix}
U \sigma_A U^\top + V & U \sigma_{AB} \\
\sigma_{AB}^\top U^\top & \sigma_B
\end{pmatrix}. 
}

\section{Quantum optimal transport for Gaussian States}
\label{sec:QOTforGaussian}

\subsection{Problem Statement}

Within the family of Gaussian states, our goal is to minimize the cost of the coupling in order to obtain the Wasserstein distance between two Gaussian states.
In this scenario, it is important to note that the minimization 
%in $D(\rho_A,\rho_B)^2$ 
can be restricted to  Gaussian couplings only, because the cost itself is quadratic in the quadratures \cite{DePalma2021}. This allows the problem to be solved entirely within the Gaussian framework, that is at the level of displacement operators and covariance matrices. In particular, it means that the coupling $\Pi$ between the Gaussian states $\rho_A$ and $\rho_B$, characterised respectively by covariance matrices $A$ and $B$ can be entirely described by its covariance matrix 
\eq{\gamma_\Pi=\begin{pmatrix}
    A&X\\X^\top&B
\end{pmatrix}\label{CouplingCovarianceMatrix}.}
Because $\rho_A$ and $\rho_B$  are physical states, their covariance matrices are necessarily symmetric and real and need to satisfy the uncertainty principle \eqref{Omega}. 
To represent a physical state, the covariance matrix $\gamma_\Pi$ must also satisfy the uncertainty principle, meaning that
\eq{\gamma_\Pi\pm 
\frac{i  }{2}\begin{pmatrix}
    \Omega&0\\0&-\Omega
\end{pmatrix}
\geq 0.
\label{CouplingPhysical}}
Note here that there is a minus sign in front of the second $\Omega$ matrix, due to the definition of the commutation relation in $\mathcal{H}^*$. 

 To find the Wasserstein distance, we need to minimize $\tr[C\Pi]$ where $C$ is given in Eq.~\eqref{cost} and $\Pi$ has the covariance $\gamma_\Pi$ given in Eq.~\eqref{CouplingCovarianceMatrix}. Using the covariance matrix, the cost can be written as
\eqarray{\label{cost in terms of covariance}C(\Pi)&=&\tr[C\Pi]=\frac12\sum_i\tr[(\hat R_i\otimes\mathds{1}-\mathds{1}\otimes \hat R_i^\top)\Pi(\hat R_i\otimes\mathds{1}-\mathds{1}\otimes \hat R_i^\top)]\nonumber\\
&=&\frac12\sum_i\Big((\gamma_\Pi)_{i,i}- 2(\gamma_\Pi)_{i,n+i}+(\gamma_\Pi)_{n+i,n+i}\Big)\nonumber\\
&=&\frac12\tr[A+B-2X]=\frac12\tr[\gamma_\Pi]-\tr[X]}
Since $A$ and $B$ are given, the Wasserstein distance is obtained by finding the optimal $X$ that minimizes the cost. 
In this paper,  we center all states at the origin for simplicity, but taking the mean vectors into account is easy. We show in Appendix~\ref{sec:noncentered} that it suffices to add to the Wasserstein distance of centered Gaussian states the norm of the difference of the mean vectors. 

We mentioned that the physicality of the coupling can be verified with the uncertainty principle, but learning from \cite{DePalma2024} that the quantum optimal plan is a quantum channel, we can force the transformation between the Gaussian states $\rho_A$ and $\rho_B$ to be itself a Gaussian  CP-map. 
Remember that the coupling can be written as in Eq.~\eqref{CouplingAsChannel}
 where $\ket{\ket{\sqrt{\rho_B}}\!} \in \mathcal{H}\otimes\mathcal{H}^*$ is the canonical purification of $\rho_B$. %\red{Not sure if it's better to use $\ket{\Psi}$ or $\|\sqrt{\rho_B}\rangle\rangle$} 
For Gaussian states, taking a canonical purification of $\rho_B$ means that both marginals have $B$ as covariance matrix and the 2$n$-mode covariance matrix of the purification $\|\sqrt{\rho_B}\rangle\!\rangle$ is \footnote{Note that because we apply a purification on the space $\mathcal{H}\otimes\mathcal{H}^*$ the off-diagonal matrix $F$ is not multiplied by the Pauli matrix $\begin{pmatrix}
    1&0\\0&-1
\end{pmatrix}$ as usual, that is like in an EPR state \cite{weedbrook, HolevoPurification}.}
\eqarray{\gamma_{\|\sqrt{\rho_B}\rangle\rangle}=\begin{pmatrix}
    B&F\\F^\top&B
\end{pmatrix}\qquad
\text{with}\ \ \ F=S\left(\bigoplus_{k=1}^n\sqrt{\nu_k^2-\frac14}\mathds{1}\right)S^\top,}
where $\nu_k$ are the symplectic values of $B$ and $S$ is the symplectic matrix that transforms $B$ in its Williamson form: $B=S\left(\bigoplus_{k=1}^n\nu_k\mathds{1}\right)S^\top$.
Therefore, following Section \ref{Gaussian channels}, the above matrix acts as the input of the CP-map and  the (output) coupling matrix can be written as
\eq{\gamma_{\Pi}=\begin{pmatrix}
    A & X\\ X^\top&B
\end{pmatrix}=\begin{pmatrix}
    UBU^\top+V&UF\\F^\top U^\top&B
\end{pmatrix}\label{CouplingAsChannelOutput}}
where $U$ and $V$ matrices have to respect the constraint \eqref{ChannelConstraint}.

\subsection{Wasserstein distance between thermal states}
\label{sec:Wasserstein Thermal states}
Before diving into the general formula for any Gaussian states, let us look at the only example known until now (to our knowledge) derived in \cite{DePalma2024}, where they analytically computed the Wasserstein distance between two one-mode thermal states $\rho_A$ and $\rho_B$ with displacement vectors $\boldsymbol d_A=\boldsymbol d_B=0$ and covariance matrices $A=\nu_A\mathds{1}$ and $B=\nu_B\mathds{1}$. 
According to Eq.\eqref{cost in terms of covariance}, the cost reduces to $\frac12\tr[\gamma_\Pi]-\tr[X]$,
and because of the shape of the covariance matrices of thermal states, we can assume we are looking for a diagonal matrix $X'=\frac{\tr X}{2}\mathds{1}=x\mathds{1}$ (see Appendix \ref{Appendix Shape of X} for more details). Therefore, we write
\eq{\gamma_\Pi=\begin{pmatrix}
    \nu_A&0&x&0\\
    0&\nu_A&0&x\\
    x&0&\nu_B&0\\
    0&x&0&\nu_B
\end{pmatrix}.}
Satisfying the uncertainty principle \eqref{CouplingPhysical} implies that the eigenvalues of this expression have to be positive which means that $|x|\leq\sqrt{\left(\nu_A-\frac{1  }{2}\right) \left(\nu_B+\frac{1  }{2}\right)}$ and $|x|\leq\sqrt{\left(\nu_B-\frac{1  }{2}\right) \left(\nu_A+\frac{1  }{2}\right)}$. 
Assuming  $\nu_A\geq\nu_B\geq\frac12$, we can simply use the constraint $|x|\leq \sqrt{\left(\nu_A+\frac{1  }{2}\right) \left(\nu_B-\frac{1  }{2}\right)}$ (the other one being automatically satisfied)
and the minimum cost is  given by 
\eqarray{D(\rho_A,\rho_B)^2
=\nu_A+\nu_B -2\sqrt{\left(\nu_A+\frac{1}{2}\right) \left(\nu_B-\frac{1}{2}\right)}
=\left(\sqrt{\nu_A+\frac12}-\sqrt{\nu_B-\frac12}\right)^2
\label{eq:WassersteinThermalDepalma}
}
It was also shown in \cite{DePalma2021} that the optimal transport plan, i.e. the optimal channel, between two thermal states turns out to be the noiseless quantum Gaussian attenuator or amplifier, with the specific choice determined by the ordering of the symplectic values $\nu_A$ and $\nu_B$.

Note  that Eq.~\eqref{eq:WassersteinThermalDepalma} can be rewritten in terms of the mean photon number $\mean{\hat n}=\nu-\frac12$ and we get
\begin{equation}
D(\rho_A,\rho_B)^2=\mean{\hat n}_A+\mean{\hat n}_B+1-2\sqrt{\mean{\hat n}_B(\mean{\hat n}_A+1)}\leq \mean{\hat n}_A+\mean{\hat n}_B+1
\end{equation}
where $\mean{\hat n}_A+\mean{\hat n}_B+1$ is the total noise (sum of the variances) \cite{Yadin18,Kwon2} of the coupling.

\subsection{Wasserstein distance between Gaussian states related by a unitary transformation}
\label{sec:unitaryrelated}

Another interesting family of states is the one of all Gaussian states related by a unitary transformation that is
\eq{\rho_A=\mathcal{U}_\phi(\rho_B)=\e^{-i\phi \hat O}\rho_B \e^{i\phi \hat O}}
with $\hat O$ being any observable. Those families have been studied for other distances such as the relative Rényi entropies \cite{Pires} and in the context of state discrimination \cite{Acevedo}.
Because we focus here only on Gaussian states, it is equivalent to say that the covariance matrices $A$ and $B$ are related by a symplectic transformation.  A unitary transformation being a special case of a Gaussian channel, it implies that $A$ can be written in terms of $B$ as in Eq.~\eqref{defchannel}, where $U$ is a symplectic matrix and $V=0$. 
According to Eq.~\eqref{CouplingAsChannelOutput}, the coupling covariance matrix is then given by 
\eq{\gamma_\Pi=\begin{pmatrix} UBU^\top &UF\\F^\top U^\top&B\end{pmatrix}.}
It is important to note that $U$ is fully determined by $A=UBU^\top$ and therefore, there is no optimization that needs to be done to compute the optimal cost. The Wasserstein distance is directly given by 
\eq{D^2(\mathcal{U}_\phi(\rho_B),\rho_B)=\frac12\tr[UBU^\top+B-2UF].}

If $\rho_B=\rho_{\rm{th}}$ is a thermal state with $B=\nu_B\mathds{1}=\frac12\frac{1+q}{1-q}$ 
(so $S=\mathds{1}$ since the covariance matrix is already in its Williamson form)
and $U$ is the identity matrix, then we are back at computing the self-distance of $\rho_B$:
\eq{D^2(\rho_{\rm{th}},\rho_{\rm{th}})=\tr[B-\sqrt{\nu_B^2-1/4}\mathds{1}]=\frac{1-\sqrt{q}}{1+\sqrt{q}},}
which coincides with Eq.~\eqref{eq:WassersteinThermalDepalma}.
If $U$ is again the identity matrix, but $\rho_B=\rho_{\rm{sqth}}$ is a squeezed thermal state (see Eq.~\eqref{gammasqth}) with 
    $\nu_B=\frac12\frac{1+q}{1-q}$ and  $S=\begin{psmallmatrix}
    \e^r&0\\0&\e^{-r}\end{psmallmatrix}$ then the self-distance of is 
\eq{D^2(\rho_{\rm{sqth}},\rho_{\rm{sqth}})=\frac{1-\sqrt{q}}{1+\sqrt{q}}\cosh(2r).}
In both cases, we recover the expected relationship between the self-distance and  the Wigner-Yanase skew information~\eqref{SkewInfoWitness}, $D^2(\rho_B,\rho_B)=2\mathcal{N}(\rho_B)$ [cf. Eq.(25) and (27) in \cite{WignerYanase}].

\section{Wasserstein distance between any two Gaussian states}
\label{sec:Wasserstein Gaussian states}
\subsection{Proof of the main formula}

We now want to extend the computation of the Wasserstein distance to any Gaussian states (still centered at the origin) and thus prove our main result given in Theorem~\ref{thm:Mainresult}. This distance will be given by the minimal cost given in Eq.~\eqref{cost in terms of covariance} where $\gamma_\Pi$ is the covariance matrix \eqref{CouplingAsChannelOutput}. In other words, we have to find the matrix $U$ that maximizes \eq{\tr[X]=\tr[UF]=\tr[U S\left(\bigoplus_{k=1}^n\sqrt{\nu_k^2-\frac14}\mathds{1}\right)S^\top]} such that 
\eq{V \pm   \frac{i}{2}\Omega \mp   \frac{i}{2} U \Omega U^\top
\geq 0 \quad\Leftrightarrow\quad A-UBU^\top \pm    \frac{i}{2}\Omega \mp  \frac{i}{2} U \Omega U^\top\geq 0 . %\quad\Leftrightarrow\quad A^\pm -UB^\pm U^\top\geq0
\label{channelconstraint}}
Note that if $A=UBU^\top$, it means the channel is a unitary transformation ($V=0$) and this case has been studied separately in Section \ref{sec:unitaryrelated}.

From now on, we will only consider covariance matrices of one-mode state, that is $2\times 2$ matrices. Therefore, we have $\det A=\nu_A^2$, $\det B=\nu_B^2$, $B=\nu_B SS^\top$, $F=\sqrt{\nu_B^2-\frac14}SS^\top=\frac{\sqrt{\nu_B^2-\frac14}}{\nu_B}B$, and $ X=UF=\frac{\sqrt{\nu_B^2-\frac14}}{\nu_B}UB$.
Since the matrix $A-UBU^\top + \frac{i}{2}\Omega -\frac{i}{2} U \Omega U^\top$ is hermitian, Eq.~\eqref{channelconstraint} implies that the real part alone is positive (see Lemma~\ref{lemma:Mpositive} in Appendix~\ref{Appendix:Lemmas}), that is 
\begin{align}
    A\geq UBU^\top\!
    \iff\sqrt{B}A\sqrt{B}\geq\sqrt{B}U\sqrt{B}\sqrt{B}U^\top\!\sqrt{B}
      \iff N^\top\! N\geq (\sqrt{B}U^\top\!\sqrt{B})^\top\!(\sqrt{B}U^\top\!\sqrt{B})
\end{align}
where $N=\sqrt{\sqrt{B}A\sqrt{B}}$.
According to the Douglas factorization Lemma (see Lemma~\ref{lemma:Douglas} in Appendix~\ref{Appendix:Lemmas}), it means that there exists a matrix $W$ with $W^\top W\leq\mathds 1$ such that $\sqrt{B}U^\top\sqrt{B}=WN$, that is
\begin{align}
 W=\sqrt{B}U^\top\sqrt{B}N^{-1} &\quad\Leftrightarrow\  \quad U^\top=\sqrt{B^{-1}}WN\sqrt{B^{-1}}\label{eqforU}
      \\&\quad \Leftrightarrow\quad  X=\frac{\sqrt{\nu_B^2-\frac14}}{\nu_B}\sqrt{B^{-1}}NW^\top\sqrt{B}. \label{eqforX}
\end{align}
The problem now reduces to 
\eq{\max\tr[X]=\frac{\sqrt{\nu_B^2-\frac14}}{\nu_B}\max\tr[NW^\top]=\frac{\sqrt{\nu_B^2-\frac14}}{\nu_B}\max\tr[WN]\label{optimizationProblem}}
such that $W^\top W\leq \mathds 1$. But that is not the only constraint on $W$. Indeed, going back to 
Eq.~\eqref{channelconstraint}, replacing $U$ by Eq.~\eqref{eqforU} and multiplying each side first by $\sqrt{B}$, then by $N^{-1}$, we have
\begin{align}
W^\top\left(\mathds{1}\pm  \frac{i}{2}\sqrt{B^{-1}} \Omega \sqrt{B^{-1}}\right) W &\leq\mathds{1}\pm  \frac{i}{2}(N^{-1}\sqrt{B}\Omega\sqrt{B} N^{-1})\nonumber\\
\iff\qquad W^\top\left(\mathds{1}\pm  \frac{i}{2\sqrt{\det B}} \Omega \right) W &\leq\mathds{1}\pm  \frac{i\sqrt{\det B}}{2\det N}\Omega\nonumber\\ 
\iff\qquad\qquad W^\top\left(\mathds{1}\pm  \frac{i}{2\nu_B} \Omega \right) W& \leq\mathds{1}\pm  \frac{i}{2\nu_A}\Omega,
\label{secondConstraintW}
\end{align}
where we used the fact that for any  $2\times2$ matrix $M$, we have 
$M\Omega M^\top= (\det M ) \ \Omega $. 

According to Lemma.~\ref{Lemma:optimizationOverPostiveW}, we know that the optimal $W$ will be positive semi-definite and symmetric and thus, using Lemma~\ref{LEM-2b2-GEN-FORM}, maximizing $\tr[W]$, under the constraint \eqref{secondConstraintW}, has the solution
\eq{W^*=\left\{\begin{array}{ccc}
          \sqrt{\frac{2+\frac1{\nu_A}}{2+\frac1{\nu_B}}}\mathds{1} & \text{if }& \nu_A\geq \nu_B \\
     \sqrt{\frac{2-\frac1{\nu_A}}{2-\frac1{\nu_B}}}\mathds{1} & \text{if }& \nu_A\leq \nu_B
    \end{array}\right.=\left\{\begin{array}{ccc}
          \sqrt{\frac{\nu_B}{\nu_A}}\sqrt{\frac{2\nu_A+1}{2\nu_B+1}}\mathds{1} & \text{if }& \nu_A\geq \nu_B \\
          &&\\
   \sqrt{\frac{\nu_B}{\nu_A}}\sqrt{\frac{2\nu_A-1}{2\nu_B-1}}\mathds{1} & \text{if }& \nu_A\leq \nu_B.
    \end{array}\right
.\label{optimalW}}
One can easily verify that both, when $\nu_A\leq\nu_B$ and $\nu_A\geq\nu_B$, $W^\top W\leq\mathds1$.
Finally, using Lemma~\ref{LEM-SOL-SOL}, we know that the maximum of $\tr[WN]$ under the same constraint \eqref{secondConstraintW}, will also be given by $W^*$ in Eq.~\eqref{optimalW}. 
Looking back at Eq.~\eqref{eqforX}, the optimal matrix $X^*$ is given by 
\begin{align}
X^*\!&=\frac{\sqrt{\nu_B^2-\frac14}}{\nu_B}\sqrt{B^{-1}}N(W^*)^\top\!\sqrt{B}
    =\!\left\{\!\begin{array}{ccc}
  \frac{\sqrt{2\nu_B-1}\sqrt{2\nu_A+1}}{2\sqrt{\nu_A\nu_B}}\sqrt{B^{-1}}\sqrt{\!\sqrt{B}A\sqrt{B}}\sqrt{B}& \text{if}& \nu_A\geq \nu_B \\
          &&\\
\frac{\sqrt{2\nu_B+1}\sqrt{2\nu_A-1}}{2\sqrt{\nu_A\nu_B}}\sqrt{B^{-1}}\sqrt{\!\sqrt{B}A\sqrt{B}}\sqrt{B} & \text{if}& \nu_A\leq \nu_B
    \end{array}\right.\label{solutionOptimlaX}
\end{align}
and we verified in Appendix~\ref{Appendix:PhysicalityOptimalX} that the solution yields a physical coupling, hence representing a valid quantum channel. We can now compute the 
the Wasserstein distance  between two Gaussian states as $\frac12\tr[\gamma_\Pi]-\tr[X^*]$ to obtain 
\begin{align}
\label{WassersteinDist main result}
D^2(\rho_A,\rho_B)&=
\left\{\begin{array}{ccc}
 \frac12\tr[A+B]-\frac12\sqrt{\frac{(2\nu_B-1)(2\nu_A+1)}{\nu_A\nu_B}}\tr[\sqrt{\sqrt{B}A\sqrt{B}}]& \text{if }& \nu_A\geq \nu_B \\
          &&\\
\frac12\tr[A+B]-\frac12\sqrt{\frac{(2\nu_B+1)(2\nu_A-1)}{\nu_A\nu_B}}\tr[\sqrt{\sqrt{B}A\sqrt{B}}] & \text{if }& \nu_A\leq \nu_B.
    \end{array}\right.\\
    &\nonumber\\
    &= \frac12\tr[A+B]-\frac12\sqrt{\frac{4\sqrt{\det A}\sqrt{\det B}-2|\sqrt{\det A}\!-\!\sqrt{\det B}|-1
 }{\sqrt{\det A}\sqrt{\det B}}}\tr[\sqrt{\!\sqrt{B}A\sqrt{B}}].\nonumber
\end{align}
Note that the formula is also valid if one or both states are pure states. Indeed, pure Gaussian states are the only Gaussian states minimizing the uncertainty principle \cite{serafini}. This means that if $\rho_A$ is pure, then $\nu_A=\frac12$  and thus, the formula to compute the Wasserstein distance simplifies to 
$D(\ket{\psi},\rho_B)^2=\frac12\tr [A+ B]$.
This is coherent with Eq.~\eqref{WassersteinPureStates} since $\tr[A]=\frac12(\Delta^2(\hat X)_\psi+\Delta^2(\hat P)_\psi)=\mean{\hat n}_\psi+\frac12,$
and similarly for $B$. The same is true if we choose $\rho_B$ to be pure.

\subsection{Examples}

If $A=\nu_A\mathds 1$ and $B=\nu_B\mathds 1$ are thermal states, with $\nu_A\geq\nu_B\geq\frac12$, then $\sqrt{\sqrt{B}A\sqrt{B}}=\sqrt{\nu_B\nu_A}\mathds{1}$
and one can check that the Wasserstein distance \eqref{WassersteinDist main result} is equal to Eq.~\eqref{eq:WassersteinThermalDepalma}.

If $A=\nu_A\begin{psmallmatrix}
    \e^{-2r_A}&0\\0&\e^{2r_A}\end{psmallmatrix}$, $B=\nu_B\begin{psmallmatrix}
    \e^{-2r_B}&0\\0&\e^{2r_B}\end{psmallmatrix}$ are squeezed thermal states (with $\nu_A\geq\nu_B\geq\frac12$ and $r_A,r_B\geq0$), then $  \sqrt{\!\sqrt{B}A\sqrt{B}} =\sqrt{\nu_B\nu_A}\begin{psmallmatrix}
            \e^{-r_A-r_B}&0\\0&\e^{r_A+r_B}
    \end{psmallmatrix}$
and the Wasserstein distance is given by    
\eq{
D^2(\rho_A,\rho_B)=\nu_A\cosh(2r_A)+\nu_B\cosh(2r_B)-\sqrt{(2\nu_B-1)(2\nu_A+1)}\cosh(r_A+r_B).
}

\subsection{Properties}
Some important properties of the Wasserstein distance between Gaussian states can be easily derived:
\begin{enumerate}
    \item $D^2(\rho_A,\rho_B)$ is symmetric under the exchange of $\rho_A$ and $\rho_B$.
    \item $D^2(\rho_A,\rho_B)$ is invariant under global rotation of the system (same rotation applied to both matrices).
    \item $D^2(\rho_A,\rho_B)$ is jointly convex, that is
    \eq{D^2(\eta \rho_{A_1}+(1-\eta)\rho_{A_2},\eta \rho_{B_1}+(1-\eta)\rho_{B_2})\leq \eta D^2(\rho_{A_1},\rho_{B_1})+(1-\eta)D^2(\rho_{A_2},\rho_{B_2}).}
    \item When looking at the classical limit, that is when $\hbar\rightarrow0$, we recover the classical formula of the Wasserstein distance between two Gaussian probability distributions.
\end{enumerate}

\begin{proof}
\begin{enumerate}
\item This is directly seen from Eq.~\eqref{WassersteinDist main result}.
\item Let $R(\theta)$ be a rotation, such that $R^\top(\theta)R(\theta)=\mathds{1}$, then
 \eq{\tilde{\gamma}_\Pi=\begin{pmatrix}
   R(\theta) & 0\\ 0&R(\theta)
\end{pmatrix}\gamma_{\Pi}\begin{pmatrix}
   R^\top(\theta) & 0\\ 0&R^\top(\theta)
\end{pmatrix}=\begin{pmatrix}
    R(\theta)AR^\top(\theta) & R(\theta)XR^\top(\theta)\\ (R(\theta)XR^\top(\theta))^\top&R(\theta)BR^\top(\theta)
\end{pmatrix}}
and, according to Eq. \eqref{cost in terms of covariance}, the cost becomes 
\eq{\tr[C\tilde{\gamma}_\Pi]=\frac{1}{2}\tr [R(\theta)(A+B-2X)R^\top(\theta)]=\frac{1}{2}\tr [A+B-2X]=\tr[C\gamma_\Pi].}
\item The joint convexity can be proven using Theorem 4 in \cite{DePalma2021} where they showed that the Wasserstein distance is convex under beam splitter mixing. Indeed, as a result of mixing two Gaussian states on a beam splitter (and tracing out the second mode), the output state is a Gaussian state with covariance matrix $A_\eta=\eta A_1 + (1-\eta)A_2$.
\item  One can re-introduce the $\hbar$ in the formula for the Wasserstein distance\footnote{It means that the uncertainty principle now read $\gamma\pm\frac{i\hbar}{2}\Omega\geq0$ or else, for all the symplectic values, $\nu\geq\frac{\hbar}{2}$.} and obtain
\eq{\label{WassersteinDist with hbar}
D^2(\rho_A,\rho_B)\!=\!
\left\{\!\begin{array}{ccc}
 \frac12\tr[A\!+\!B]-\frac12\sqrt{\frac{(2\nu_B-\hbar)(2\nu_A+\hbar)}{\nu_A\nu_B}}\tr[\sqrt{\!\sqrt{B}A\sqrt{B}}]& \text{if}& \nu_A\geq \nu_B \\
          &&\\
\frac12\tr[A\!+\!B]-\frac12\sqrt{\frac{(2\nu_B+\hbar)(2\nu_A-\hbar)}{\nu_A\nu_B}}\tr[\sqrt{\!\sqrt{B}A\sqrt{B}}] & \text{if}& \nu_A\leq \nu_B.
    \end{array}\right.
}
At the classical limit, that is when $\hbar\rightarrow0$, we recover, as expected, the classical formula of the Wasserstein distance between two Gaussian probability distributions \cite{givens84}, \eq{W_2^2(\mathcal{N}(0, \Sigma_1), \mathcal{N}(0, \Sigma_2)) = \frac{1}{2}\text{Tr} \left( \Sigma_1 + \Sigma_2 - 2\left( \Sigma_1^{1/2} \Sigma_2 \Sigma_1^{1/2} \right)^{1/2} \right).}
\end{enumerate}
\end{proof}

\section{Comparison to other distances}
\label{sec:distances}

Distance measures can take operational meaning in various contexts and can potentially be embedded within a resource-theoretic framework. For example, it was recently established that the Wigner-Yanase skew information serves as a valid resource measure in the resource theory of asymmetry \cite{Takagi2019Skew}, which is particularly relevant in the context of quantum metrology. Given that the Wasserstein self-distance coincides with the Wigner-Yanase skew information, this suggests that one may extend the properties from this resource theory to the Wasserstein distance in general, which could lead to a new operational interpretation. On the other hand, distance and similarity measures such as fidelity, trace distance, or the Bures metric are widely used to quantify the distinguishability of quantum states. The role of these measures is particularly significant as distinguishing quantum states~\cite{NielsenChuang2000,Helstrom1,Bae_2015} is a fundamental and unavoidable task in quantum information theory with implications in quantum hypothesis testing \cite{Yuen}, metrology \cite{paris2009quantumest}, quantum sensing~\cite{Hayashi2006}, and secure quantum communication~\cite{RevModPhys.74.145}.

For classical probability distributions, a detailed overview of relationships between various metrics and distances and how they bound each other can be found in \cite{Gibbs2002}. Of particular interest is the connection between the Wasserstein metric and the total variation distance, which is the classical equivalent to the trace distance. Given the computational complexity of the trace distance and its significance as a measure with direct operational interpretation in state discrimination, any potential connection between the Wasserstein distance and the trace distance will be of interest and will shed new light on recent progress in bounding the trace distance \cite{Holevo2025TraceNorm,Mele2025LearningCV}. We anticipate that an equivalent relation may hold for the quantum case, which will be the subject of future work. In this section, we explicitly compare the Wasserstein distance with various measures having a closed-form solution for Gaussian states and provide evidence for an inequality between the Wasserstein distance and the Bures distance.

In Table~\ref{tab:comparisonDistancesForThermal}, we compare the Wasserstein distance to the fidelity, the Bures distance, the relative entropy and the Hilbert-Schmidt distance (all properly defined in Appendix \ref{Appendix:defDistances})
for general one-mode (centered) Gaussian states and also give the simpler formula in the case of thermal states with covariance matrices $A=(\nu+\theta)\mathds{1}$, $B=\nu\mathds{1}$. 
  \begin{table*}[h!]
    \centering
    \footnotesize
    \setlength{\tabcolsep}{1pt} % default is 6pt
\renewcommand{\arraystretch}{2} % Increases row height (default is 1.0)
    % \begin{tabular}{|c|p{10.9cm}|}
        \begin{tabular}{|c|c|c|}
     \hline
\cellcolor{quantumviolet!5!} Distances & \cellcolor{quantumviolet!5!}  For Gaussian states  & \cellcolor{quantumviolet!5!} For thermal states  \\
    \hline
\makecell{Quantum\\ Wasserstein\\$D_W(\rho_A,\rho_B)$} & Eq.~\eqref{WassersteinDist main result} & $\sqrt{\nu+\theta+\frac12}-\sqrt{\nu-\frac12}$
\\
\hline
    \makecell{Fidelity \\$F(\rho_A,\rho_B)$}& 
   \scriptsize $\begin{array}{l}
            \left(\sqrt{\det(A\!+\!B)+4\left(\det A\!-\!\frac14\right)\left(\det B\!-\!\frac14\right)}\right.\\\left.-\sqrt{4\left(\det A-\frac14\right)\left(\det B-\frac14\right)}\right)^{-1}\end{array}$
        &
        \normalsize $\frac{2}{4(\nu^2+\nu\theta)+1-\sqrt{\left(4\nu^2-1\right)\left(4(\nu+\theta)^2-1\right)}}$\\  
        \hline
        \makecell{Bures distance\\$D_{Bures}(\rho_A,\rho_B)$} &$\sqrt{2-2\sqrt{F(\rho_A,\rho_B)}}$ &
 $\begin{array}{l}
           \sqrt{ 2-2\sqrt{\frac{2}{4(\nu^2+\nu\theta)+1-\sqrt{\left(4\nu^2-1\right)\left(4(\nu+\theta)^2-1\right)}}}}\\=
               \sqrt{2-\frac{\sqrt{2}\sqrt{4(\nu^2+\nu\theta)+1+\sqrt{\left(4\nu^2-1\right)\left(4(\nu+\theta)^2-1\right)}}}{2\nu+\theta}  }
            \end{array}$
        \\
        \hline
 \makecell{ Overlap\\$\mathcal{O}(\rho_A,\rho_B)$} & $\left(\sqrt{\det(A+B)}\right)^{-1}$& \normalsize$\frac1{2\nu+\theta}$   \\
        \hline
       \makecell{ Hilbert\\-Schmidt  \\$D_{HS}(\rho_A,\rho_B)$}& 
       $\sqrt{\frac1{\sqrt{2\det[A]}}-2\frac1{\sqrt{\det[A+B]}}+\frac1{\sqrt{2\det[B]}}}$&\normalsize$\frac{\theta}{\sqrt{2\nu( \nu+\theta)  (2\nu+\theta)}}$ \\
        \hline
\makecell{Relative\\ entropy %\cite{Parthasarathy,Seshadreesan}
\\$S(\rho_A \| \rho_B)$} & 
$\begin{array}{l}
        \left(\sqrt{\det[A]}+\frac12\right)\log\left(\frac{\sqrt{\det[B]}+\frac12}{\sqrt{\det[A]}+\frac12}\right)  \\
            + \left(\sqrt{\det[A]}-\frac12\right)\log\left(\frac{\sqrt{\det[A]}-\frac12}{\sqrt{\det[B]}-\frac12}\right) \\
        + \frac12\left(\tr \tilde A\!-\!2\sqrt{\det[A]}\right)\log\left(\frac{\sqrt{\det[B]}+\frac12}{\sqrt{\det[B]}-\frac12}\right)\\
            \text{where $\tilde A$ = $S^\top AS$ and $S$ is such that}\\
SBS^\top = \nu_B.
        \end{array}$
        &$\begin{array}{c}
              \left(\nu+\theta+\frac12\right)\log\left(\frac{\nu+\frac12}{\nu+\theta+\frac12}\right)\\+\left(\nu+\theta-\frac12\right)\log\left(\frac{\nu+\theta-\frac12}{\nu-\frac12}\right)
                \end{array} $\\
        \hline
    \end{tabular}
    \caption{Comparison of distances between one-mode Gaussian states centered on $0$ with covariance matrices $A$ and $B$. The thermal states have covariance matrices $A=(\nu+\theta)\mathds{1}$ and $B=\nu\mathds{1}$}
    \label{tab:comparisonDistancesForThermal}
\end{table*}
\normalsize

\begin{figure}[h!]
    \centering
    \includegraphics[width=1\linewidth,trim=5 0 70 0,clip]{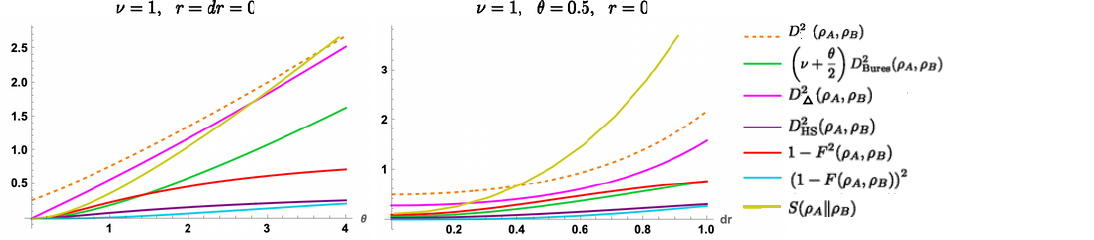}
    \caption{Comparison of distances for two squeezed thermal states $\rho_A$ and $\rho_B$ with covariance matrices $A= (\nu+\theta)\begin{psmallmatrix}
   \e^{-2r}&0\\0&    \e^{2r}
\end{psmallmatrix}$, $B=\nu\begin{psmallmatrix}
   \e^{-2(r+dr)}&0\\0&    \e^{2(r+dr)}
\end{psmallmatrix}$. The labels of the curves refer to the quantities in Table~\ref{tab:comparisonDistancesForThermal}.}
    \label{Plots}
\end{figure}

 In Figure~\ref{Plots} we show how those distances behave for two squeezed thermal states $\rho_A$ and $\rho_B$ with covariance matrices $A= (\nu+\theta)\begin{psmallmatrix}
   \e^{-2r}&0\\0&   \e^{2r}
\end{psmallmatrix}$, $B=\nu\begin{psmallmatrix}  \e^{-2(r+dr)}&0\\0&    \e^{2(r+dr)}\end{psmallmatrix}$. 
One can numerically see that the Wasserstein distance is bounded by 
the Bures distances as \eq{
\frac12(\nu_A+\nu_B)D^2_{\rm{Bures}}(\rho_A,\rho_B)\leq D^2(\rho_A,\rho_B).\label{eq:inewBuresWasserstein} }
We prove it analytically for thermal states in Appendix~\ref{Appendix:ineqaulityBuresWasserstein}.
A more in-depth comparison of the distances and how they can bound each other is left for future work.

By observing the left plot, when $\theta=0$, $\rho_A=\rho_B$ and one notices that most distances start as zero, contrary to the Wasserstein distance. We thus also plotted the distance
\eq{D^2_{\Delta}(\rho_A,\rho_B)=
D^2(\rho_A,\rho_B)-\frac12D^2(\rho_A,\rho_A)-\frac12D^2(\rho_B,\rho_B).} 
This quantity was introduced in \cite{DePalma2024} and might be a better definition of a distance, since it indeed gives 0 between a state and itself. In particular, the authors conjectured that this quantity satisfies the triangle inequality, a conjecture that was proven under certain conditions in \cite{Bunth20214} and \cite{BunthPitrikTitkosVirosztek2025}.

\section{Conclusion}
\label{sec:conclusion}
We have provided an explicit, fully analytical expression for the quantum Wasserstein distance between arbitrary one-mode Gaussian states within the transport framework where optimal transport plans are quantum channels. 
We have shown that the Wasserstein distance is symmetric and jointly convex, and that it recovers both the known formula for thermal states and the classical Wasserstein distance for Gaussian distributions in the classical limit. We also observed that it bounds the Bures distance.
This opens the door to comparing the Wasserstein distance with other well-known distances between quantum states, thereby providing an operational interpretation to this measure. In particular, the focus should be on the trace distance which captures exactly how different two quantum states can be in any physically allowed experiment, but is known to be hard to compute.

It would also be relevant to extend the formula to multi-mode Gaussian states. If the $n$-mode Gaussian state is a tensor product of one-modal Gaussian states, the Wasserstein distance can be computed as a sum of the Wasserstein distances of each mode \cite{DePalma2021}, but the formula is not known in the general case.
Another natural extension consists of treating non-Gaussian states that can be expressed as a sum of Gaussian \cite{HertzSumGaussian,XanaduSumGaussian}, a class that encompasses many relevant non-Gaussian states and appears to be a simple step beyond strict Gaussianity. 
More generally, the computation of the Wasserstein distance for non-Gaussian states remains open. In analogy with the approach of Toth et al. \cite{Toth2023quantumwasserstein}, who evaluated the Wasserstein distance over separable states and used it as an entanglement quantifier, it would be compelling to investigate whether minimization restricted to appropriate families could yield operational measures of other quantum features such as non-Gaussianity (based e.g. on the Hilbert-Schmidt distance \cite{Genoni2007}, the value of the Wigner function at the origin \cite{Genoni2013}, the quantum relative entropy \cite{GenoniParis2008}, the stellar rank \cite{Chabaud2020}), or nonclassicality (e.g. the Wigner negativity \cite{Kenfack_2004}, the quadrature coherence scale \cite{Debievre,Hertz}, inequalities with other quasi-probability distributions \cite{Agudeloetal2013,Ryl2,Bohmann}).

\begin{acknowledgments}
AH and KH acknowledge that the NRC headquarters is located on the traditional unceded territory of the Algonquin Anishinaabe and Mohawk people, as well as support from NRC's Applied Quantum Computing Challenge Program. KH and AG acknowledge support from NSERC alliance programs. The authors thank Noah Lupu-Gladstein for fruitful conversations. 
\end{acknowledgments}

\bibliographystyle{quantum}
% \bibliography{references}

\begin{thebibliography}{10}

\bibitem{Monge1781}
Gaspard Monge.
\newblock ``M\'emoire sur la th\'eorie des d\'eblais et des remblais''.
\newblock Histoire de l'Acad\'emie Royale des Sciences de Paris~(1781).

\bibitem{Kantorovich1942}
L.~Kantorovich.
\newblock ``On the translocation of masses''.
\newblock \href{http://www.jstor.org/stable/2626967}{C. R. (Doklady) Acad. Sci. URSS (N.S.) {\bf 37}, 199--201}~(1942).

\bibitem{PeyreCuturi2019COT}
Gabriel Peyr{\'e} and Marco Cuturi.
\newblock ``Computational optimal transport''.
\newblock \href{https://dx.doi.org/10.1561/2200000073}{Foundations and Trends in Machine Learning {\bf 11}, 355--607}~(2019).

\bibitem{Caracciolo2014Scaling}
S.~Caracciolo, C.~Lucibello, G.~Parisi, and G.~Sicuro.
\newblock ``Scaling hypothesis for the euclidean bipartite matching problem''.
\newblock \href{https://dx.doi.org/10.1103/PhysRevE.90.012118}{Physical Review E {\bf 90}, 012118}~(2014).

\bibitem{Otto2001Geometry}
Felix Otto.
\newblock ``The geometry of dissipative evolution equations: The porous medium equation''.
\newblock \href{https://dx.doi.org/10.1081/PDE-100002243}{Communications in Partial Differential Equations {\bf 26}, 101--174}~(2001).

\bibitem{Villani2009}
C\'edric Villani.
\newblock ``Optimal transport: Old and new''.
\newblock \href{https://dx.doi.org/https://doi.org/10.1007/978-3-540-71050-9}{Springer}. Berlin~(2009).

\bibitem{ZyczkowskiSlomczynski1998}
K.~Zyczkowski and W.~S{\l}omczy{\'n}ski.
\newblock ``The monge distance between quantum states''.
\newblock \href{https://dx.doi.org/https://dx.doi.org/10.1088/0305-4470/31/45/009}{Journal of Physics A: Mathematical and General {\bf 31}, 9095}~(1998).

\bibitem{BianeVoiculescu2001}
P.~Biane and D.~Voiculescu.
\newblock ``A free probability analogue of the wasserstein metric on the trace-state space''.
\newblock \href{https://dx.doi.org/https://dx.doi.org/10.1007/s00039-001-8226-4}{Geometric and Functional Analysis (GAFA) {\bf 11}, 1125--1138}~(2001).

\bibitem{CarlenMaas2014}
Eric~A. Carlen and Jan Maas.
\newblock ``An analog of the 2-wasserstein metric in noncommutative probability under which the fermionic fokker--planck equation is gradient flow for the entropy''.
\newblock \href{https://dx.doi.org/https://dx.doi.org/10.1007/s00220-014-2124-8}{Communications in Mathematical Physics {\bf 331}, 887--926}~(2014).

\bibitem{CarlenMaas2017}
Eric~A. Carlen and Jan Maas.
\newblock ``Gradient flow and entropy inequalities for quantum markov semigroups with detailed balance''.
\newblock \href{https://dx.doi.org/https://dx.doi.org/10.1016/j.jfa.2017.05.003}{Journal of Functional Analysis {\bf 273}, 1810--1869}~(2017).

\bibitem{CarlenMaas2020}
Eric~A. Carlen and Jan Maas.
\newblock ``Non-commutative calculus, optimal transport and functional inequalities in dissipative quantum systems''.
\newblock \href{https://dx.doi.org/https://dx.doi.org/10.1007/s10955-019-02434-w}{Journal of Statistical Physics {\bf 178}, 319--378}~(2020).

\bibitem{RouzeDatta2019}
Cambyse Rouz{\'e} and Nilanjana Datta.
\newblock ``Concentration of quantum states from quantum functional and transportation cost inequalities''.
\newblock \href{https://dx.doi.org/https://dx.doi.org/10.1063/1.5023210}{Journal of Mathematical Physics {\bf 60}, 012202}~(2019).

\bibitem{DattaRouze2020}
Nilanjana Datta and Cambyse Rouz{\'e}.
\newblock ``Relating relative entropy, optimal transport and fisher information: A quantum {HWI} inequality''.
\newblock In Annales Henri Poincar{\'e}.
\newblock \href{https://dx.doi.org/https://dx.doi.org/10.1007/s00023-020-00891-8}{Pages 1--36}.
\newblock Springer~(2020).

\bibitem{ChenGeorgiouTannenbaum2018a}
Yongxin Chen, Tryphon~T. Georgiou, and Allen~R. Tannenbaum.
\newblock ``Matrix optimal mass transport: A quantum mechanical approach''.
\newblock \href{https://dx.doi.org/https://doi.org/10.48550/arXiv.1610.03041}{IEEE Transactions on Automatic Control {\bf 63}, 2612--2619}~(2018).

\bibitem{ChenGeorgiouTannenbaum2018b}
Yongxin Chen, Tryphon~T. Georgiou, and Allen~R. Tannenbaum.
\newblock ``Wasserstein geometry of quantum states and optimal transport of matrix-valued measures''.
\newblock In Emerging Applications of Control and Systems Theory.
\newblock Pages 139--150.
\newblock Springer~(2018).

\bibitem{GolseMouhotPaul2016}
Fran\c{c}ois Golse, Cl\'ement Mouhot, and Thierry Paul.
\newblock ``On the mean field and classical limits of quantum mechanics''.
\newblock \href{https://dx.doi.org/https://dx.doi.org/10.1007/s00220-015-2485-7}{Communications in Mathematical Physics {\bf 343}, 165--205}~(2016).

\bibitem{CagliotiGolsePaul2018preprint}
Emanuele Caglioti, Fran\c{c}ois Golse, and Thierry Paul.
\newblock ``Towards optimal transport for quantum densities''.
\newblock Working paper or preprint~(2018).

\bibitem{Golse2018}
Fran\c{c}ois Golse.
\newblock ``The quantum {N}-body problem in the mean-field and semiclassical regime''.
\newblock \href{https://dx.doi.org/https://dx.doi.org/10.1098/rsta.2017.0229}{Philosophical Transactions of the Royal Society A: Mathematical, Physical and Engineering Sciences {\bf 376}, 20170229}~(2018).

\bibitem{GolsePaul2017}
Fran\c{c}ois Golse and Thierry Paul.
\newblock ``The schr\"odinger equation in the mean-field and semiclassical regime''.
\newblock \href{https://dx.doi.org/https://dx.doi.org/10.1007/s00205-016-1031-x}{Archive for Rational Mechanics and Analysis {\bf 223}, 57--94}~(2017).

\bibitem{GolsePaul2018}
Fran\c{c}ois Golse and Thierry Paul.
\newblock ``Wave packets and the quadratic monge--kantorovich distance in quantum mechanics''.
\newblock \href{https://dx.doi.org/https://dx.doi.org/https://doi.org/10.1016/j.crma.2017.12.007}{Comptes Rendus Math\'ematique {\bf 356}, 177--197}~(2018).

\bibitem{CagliotiGolsePaul2020}
Emanuele Caglioti, Fran\c{c}ois Golse, and Thierry Paul.
\newblock ``Quantum optimal transport is cheaper''.
\newblock \href{https://dx.doi.org/https://dx.doi.org/10.1007/s10955-020-02571-7}{Journal of Statistical Physics}~(2020).

\bibitem{DePalma2021}
Giacomo De~Palma and Dario Trevisan.
\newblock ``Quantum optimal transport with quantum channels''.
\newblock \href{https://dx.doi.org/10.1007/s00023-021-01042-3}{Annales Henri Poincaré {\bf 22}, 3199--3234}~(2021).

\bibitem{DePalmaMarvianTrevisanLloyd2021}
G.~De~Palma, M.~Marvian, D.~Trevisan, and S.~Lloyd.
\newblock ``The quantum wasserstein distance of order 1''.
\newblock \href{https://doi.org/10.1109/TIT.2021.3076442}{IEEE Transactions on Information Theory {\bf 67}, 6627--6643}~(2021).


\bibitem{FriedlandEcksteinColeZyczkowski2022}
Shmuel Friedland, Micha\l\ Eckstein, Sam Cole, and Karol Zyczkowski.
\newblock ``Quantum monge--kantorovich problem and transport distance between density matrices''.
\newblock \href{https://dx.doi.org/https://dx.doi.org/10.1103/PhysRevLett.129.110402}{Physical Review Letters {\bf 129}, 110402}~(2022).

\bibitem{ColeEcksteinFriedlandZyczkowski2021}
Sam Cole, Micha\l\ Eckstein, Shmuel Friedland, and Karol Zyczkowski.
\newblock ``Quantum optimal transport''.
\newblock arXiv:2105.06922~(2021).

\bibitem{BistronEcksteinZyczkowski2023}
R.~Bistro\'n, M.~Eckstein, and K.~Zyczkowski.
\newblock ``Monotonicity of a quantum 2-wasserstein distance''.
\newblock \href{https://dx.doi.org/https://dx.doi.org/10.1088/1751-8121/acb9c8}{Journal of Physics A: Mathematical and Theoretical {\bf 56}, 095301}~(2023).

\bibitem{MaasRademacherTitkosVirosztek2024}
Jan Maas, Simone Rademacher, Tam\'as Titkos, and D\'aniel Virosztek, editors.
\newblock ``Optimal transport on quantum structures''.
\newblock \href{https://dx.doi.org/10.1007/978-3-031-50466-2}{Volume~29 of Bolyai Society Mathematical Studies, pages VIII+300}.
\newblock Springer. Cham~(2024).

\bibitem{beatty2025wasserstein}
Emily Beatty.
\newblock ``Wasserstein distances on quantum structures: an overview''~(2025).
\newblock  \href{http://arxiv.org/abs/2506.09794}{arXiv:2506.09794}.

\bibitem{DePalma2024}
Giacomo De~Palma and Dario Trevisan.
\newblock ``Quantum optimal transport: Quantum channels and qubits''.
\newblock \href{https://dx.doi.org/10.1007/978-3-031-50466-2_4}{Pages 203--239}.
\newblock Springer Nature Switzerland. Cham~(2024).

\bibitem{WignerYanase}
Shunlong Luo and Yue Zhang.
\newblock ``Quantifying nonclassicality via wigner-yanase skew information''.
\newblock \href{https://dx.doi.org/10.1103/PhysRevA.100.032116}{Phys. Rev. A {\bf 100}, 032116}~(2019).

\bibitem{Titulaer}
U.~M. Titulaer and R.~J. {G}lauber.
\newblock ``Correlation functions for coherent fields''.
\newblock \href{https://dx.doi.org/10.1103/PhysRev.140.B676}{Phys. Rev. {\bf 140}, B676--B682}~(1965).

\bibitem{Yadin18}
Benjamin Yadin, Felix~C. Binder, Jayne Thompson, Varun Narasimhachar, Mile Gu, and M.~S. Kim.
\newblock ``Operational resource theory of continuous-variable nonclassicality''.
\newblock \href{https://dx.doi.org/10.1103/PhysRevX.8.041038}{Phys. Rev. X {\bf 8}, 041038}~(2018).

\bibitem{Kwon2}
Hyukjoon Kwon, Kok~Chuan Tan, Tyler Volkoff, and Hyunseok Jeong.
\newblock ``Nonclassicality as a quantifiable resource for quantum metrology''.
\newblock \href{https://dx.doi.org/10.1103/PhysRevLett.122.040503}{Phys. Rev. Lett. {\bf 122}, 040503}~(2019).

\bibitem{Debievre}
Stephan De~Bi\`evre, Dmitri~B. Horoshko, Giuseppe Patera, and Mikhail~I. Kolobov.
\newblock ``Measuring nonclassicality of bosonic field quantum states via operator ordering sensitivity''.
\newblock \href{https://dx.doi.org/10.1103/PhysRevLett.122.080402}{Phys. Rev. Lett. {\bf 122}, 080402}~(2019).

\bibitem{Hertz}
Anaelle Hertz and Stephan De~Bi\`evre.
\newblock ``Quadrature coherence scale driven fast decoherence of bosonic quantum field states''.
\newblock \href{https://dx.doi.org/10.1103/PhysRevLett.124.090402}{Phys. Rev. Lett. {\bf 124}, 090402}~(2020).

\bibitem{PhysRevA.73.062330}
Stefano Olivares, Matteo G.~A. Paris, and Ulrik~L. Andersen.
\newblock ``Cloning of gaussian states by linear optics''.
\newblock \href{https://dx.doi.org/10.1103/PhysRevA.73.062330}{Phys. Rev. A {\bf 73}, 062330}~(2006).

\bibitem{PhysRevA.71.032336}
Hyunchul Nha and H.~J. Carmichael.
\newblock ``Distinguishing two single-mode gaussian states by homodyne detection: An information-theoretic approach''.
\newblock \href{https://dx.doi.org/10.1103/PhysRevA.71.032336}{Phys. Rev. A {\bf 71}, 032336}~(2005).

\bibitem{Parthasarathy}
K.~R. Parthasarathy.
\newblock ``A pedagogical note on the computation of relative entropy of two n-mode gaussian states''.
\newblock In Luigi Accardi, Farrukh Mukhamedov, and Ahmed Al~Rawashdeh, editors, Infinite Dimensional Analysis, Quantum Probability and Applications.
\newblock \href{https://dx.doi.org/https://doi.org/10.48550/arXiv.2102.06708}{Pages 55--72}.
\newblock Cham~(2022). Springer International Publishing.

\bibitem{Seshadreesan}
Kaushik~P. Seshadreesan, Ludovico Lami, and Mark~M. Wilde.
\newblock ``Rényi relative entropies of quantum gaussian states''.
\newblock \href{https://dx.doi.org/10.1063/1.5007167}{Journal of Mathematical Physics {\bf 59}, 072204}~(2018).

\bibitem{OZAWA2000158}
Masanao Ozawa.
\newblock ``Entanglement measures and the hilbert–schmidt distance''.
\newblock \href{https://dx.doi.org/https://doi.org/10.1016/S0375-9601(00)00171-7}{Physics Letters A {\bf 268}, 158--160}~(2000).

\bibitem{Coles}
Patrick~J. Coles, M.~Cerezo, and Lukasz Cincio.
\newblock ``Strong bound between trace distance and hilbert-schmidt distance for low-rank states''.
\newblock \href{https://dx.doi.org/10.1103/PhysRevA.100.022103}{Phys. Rev. A {\bf 100}, 022103}~(2019).

\bibitem{weedbrook}
Christian Weedbrook, Stefano Pirandola, Ra\'ul Garc\'{\i}a-Patr\'on, Nicolas~J. Cerf, Timothy~C. Ralph, Jeffrey~H. Shapiro, and Seth Lloyd.
\newblock ``{G}aussian quantum information''.
\newblock \href{https://dx.doi.org/10.1103/RevModPhys.84.621}{Rev. Mod. Phys. {\bf 84}, 621--669}~(2012).

\bibitem{Anaellethesis}
Anaelle Hertz.
\newblock ``Exploring continuous-variable entropic uncertainty relations and separability criteria in quantum phase space''.
\newblock \href{https://difusion.ulb.ac.be/vufind/Record/ULB-DIPOT:oai:dipot.ulb.ac.be:2013/267632/Details}{PhD thesis. Universit\'e libre de Bruxelles} ~(2018).


\bibitem{serafini}
Alessio Serafini.
\newblock ``Quantum continuous variables: a primer of theoretical methods''.
\newblock CRC press. ~(2023).

\bibitem{HolevoPurification}
A.~S. Holevo and R.~F. Werner.
\newblock ``Evaluating capacities of bosonic gaussian channels''.
\newblock \href{https://dx.doi.org/10.1103/PhysRevA.63.032312}{Phys. Rev. A {\bf 63}, 032312}~(2001).

\bibitem{Pires}
Diego~Paiva Pires, Augusto Smerzi, and Tommaso Macr\`{\i}.
\newblock ``Relating relative r\'enyi entropies and wigner-yanase-dyson skew information to generalized multiple quantum coherences''.
\newblock \href{https://dx.doi.org/10.1103/PhysRevA.102.012429}{Phys. Rev. A {\bf 102}, 012429}~(2020).

\bibitem{Acevedo}
Alberto Acevedo and Janek Wehr.
\newblock ``Asymptotic quantum state discrimination for mixtures of unitarily related states''~(2024).

\bibitem{givens84}
Clark~R. Givens and Rae~M. Shortt.
\newblock ``A class of {W}asserstein metrics for probability distributions.''.
\newblock \href{https://dx.doi.org/https://doi.org/10.1307/mmj/1029003026}{The Michigan Mathematical Journal {\bf 31}, 231--240}~(1984).

\bibitem{Takagi2019Skew}
Ryutaroh Takagi.
\newblock ``Skew informations from an operational view via resource theory of asymmetry''.
\newblock \href{https://dx.doi.org/10.1038/s41598-019-50279-w}{Scientific Reports {\bf 9}, 14562}~(2019).

\bibitem{NielsenChuang2000}
Michael~A. Nielsen and Isaac~L. Chuang.
\newblock ``{Quantum Computation and Quantum Information}''.
\newblock Cambridge University Press. Cambridge~(2000).

\bibitem{Helstrom1}
C.~W. Helstrom.
\newblock ``Quantum detection and estimation theory''.
\newblock Volume 123 of Mathematics in Science and Engineering.
\newblock Academic Press. New York~(1976).

\bibitem{Bae_2015}
Joonwoo Bae and Leong-Chuan Kwek.
\newblock ``Quantum state discrimination and its applications''.
\newblock \href{https://dx.doi.org/10.1088/1751-8113/48/8/083001}{Journal of Physics A: Mathematical and Theoretical {\bf 48}, 083001}~(2015).

\bibitem{Yuen}
Horace~P. Yuen, Robert~S. Kennedy, and Melvin Lax.
\newblock ``Optimum testing of multiple hypotheses in quantum detection theory''.
\newblock \href{https://dx.doi.org/10.1109/TIT.1975.1055351}{IEEE Transactions on Information Theory {\bf 21}, 125--134}~(1975).

\bibitem{paris2009quantumest}
Matteo~GA Paris.
\newblock ``Quantum estimation for quantum technology''.
\newblock \href{https://dx.doi.org/10.1142/S0219749909004839}{International Journal of Quantum Information {\bf 7}, 125--137}~(2009).

\bibitem{Hayashi2006}
Masahito Hayashi.
\newblock ``Quantum hypothesis testing and discrimination of quantum states''.
\newblock \href{https://dx.doi.org/10.1007/3-540-30266-2_4}{Pages 69--91}.
\newblock Springer Berlin Heidelberg. Berlin, Heidelberg~(2006).

\bibitem{RevModPhys.74.145}
Nicolas Gisin, Gr\'egoire Ribordy, Wolfgang Tittel, and Hugo Zbinden.
\newblock ``Quantum cryptography''.
\newblock \href{https://dx.doi.org/10.1103/RevModPhys.74.145}{Rev. Mod. Phys. {\bf 74}, 145--195}~(2002).

\bibitem{Gibbs2002}
Alison~L. Gibbs and Francis~Edward Su.
\newblock ``On choosing and bounding probability metrics''.
\newblock \href{https://dx.doi.org/None}{International Statistical Review {\bf 70}, 419--435}~(2002).

\bibitem{Holevo2025TraceNorm}
A.~S. Holevo.
\newblock ``On estimates of trace-norm distance between quantum gaussian states''.
\newblock \href{https://dx.doi.org/10.48550/arXiv.2408.11400}{Lobachevskii Journal of Mathematics {\bf 46}, 2542--2553}~(2025).

\bibitem{Mele2025LearningCV}
F.~A. Mele, A.~A. Mele, L.~Bittel, J.~Eisert, V.~Giovannetti, L.~Lami, L.~Leone, and S.~F.~E. Oliviero.
\newblock ``Learning quantum states of continuous-variable systems''.
\newblock \href{https://dx.doi.org/10.1038/s41567-025-03086-2}{Nature Physics {\bf 21}, 2002--2008}~(2025).

\bibitem{Bunth20214}
Gergely Bunth, J\'ozsef Pitrik, Tam\'as Titkos, and D\'aniel Virosztek.
\newblock ``Metric property of quantum wasserstein divergences''.
\newblock \href{https://dx.doi.org/10.1103/PhysRevA.110.022211}{Phys. Rev. A {\bf 110}, 022211}~(2024).

\bibitem{BunthPitrikTitkosVirosztek2025}
Gergely Bunth, J{\'o}zsef Pitrik, Tam{\'a}s Titkos, and D{\'a}niel Virosztek.
\newblock ``Wasserstein distances and divergences of order p by quantum channels''~(2025).
\newblock  \href{http://arxiv.org/abs/2501.08066}{arXiv:2501.08066}.

\bibitem{HertzSumGaussian}
Anaelle Hertz, Aaron~Z. Goldberg, and Khabat Heshami.
\newblock ``Quadrature coherence scale of linear combinations of gaussian functions in phase space''.
\newblock \href{https://dx.doi.org/10.1103/PhysRevA.110.012408}{Phys. Rev. A {\bf 110}, 012408}~(2024).

\bibitem{XanaduSumGaussian}
J.~Eli Bourassa, Nicol\'as Quesada, Ilan Tzitrin, Antal Sz\'ava, Theodor Isacsson, Josh Izaac, Krishna~Kumar Sabapathy, Guillaume Dauphinais, and Ish Dhand.
\newblock ``Fast simulation of bosonic qubits via gaussian functions in phase space''.
\newblock \href{https://dx.doi.org/10.1103/PRXQuantum.2.040315}{PRX Quantum {\bf 2}, 040315}~(2021).

\bibitem{Toth2023quantumwasserstein}
G{\'{e}}za T{\'{o}}th and J{\'{o}}zsef Pitrik.
\newblock ``Quantum {W}asserstein distance based on an optimization over separable states''.
\newblock \href{https://dx.doi.org/10.22331/q-2023-10-16-1143}{{Quantum} {\bf 7}, 1143}~(2023).

\bibitem{Genoni2007}
Marco~G. Genoni, Matteo G.~A. Paris, and Konrad Banaszek.
\newblock ``Measure of the non-gaussian character of a quantum state''.
\newblock \href{https://dx.doi.org/10.1103/PhysRevA.76.042327}{Phys. Rev. A {\bf 76}, 042327}~(2007).

\bibitem{Genoni2013}
Marco~G. Genoni, Mattia~L. Palma, Tommaso Tufarelli, Stefano Olivares, M.~S. Kim, and Matteo G.~A. Paris.
\newblock ``Detecting quantum non-gaussianity via the wigner function''.
\newblock \href{https://dx.doi.org/10.1103/PhysRevA.87.062104}{Phys. Rev. A {\bf 87}, 062104}~(2013).

\bibitem{GenoniParis2008}
Marco~G. Genoni, Matteo G.~A. Paris, and Konrad Banaszek.
\newblock ``Quantifying the non-gaussian character of a quantum state by quantum relative entropy''.
\newblock \href{https://dx.doi.org/10.1103/PhysRevA.78.060303}{Phys. Rev. A {\bf 78}, 060303}~(2008).

\bibitem{Chabaud2020}
Ulysse Chabaud, Damian Markham, and Fr\'ed\'eric Grosshans.
\newblock ``Stellar representation of non-gaussian quantum states''.
\newblock \href{https://dx.doi.org/10.1103/PhysRevLett.124.063605}{Phys. Rev. Lett. {\bf 124}, 063605}~(2020).

\bibitem{Kenfack_2004}
Anatole Kenfack and Karol Życzkowski.
\newblock ``Negativity of the wigner function as an indicator of non-classicality''.
\newblock \href{https://dx.doi.org/10.1088/1464-4266/6/10/003}{Journal of Optics B: Quantum and Semiclassical Optics {\bf 6}, 396}~(2004).

\bibitem{Agudeloetal2013}
E.~Agudelo, J.~Sperling, and W.~Vogel.
\newblock ``Quasiprobabilities for multipartite quantum correlations of light''.
\newblock \href{https://dx.doi.org/10.1103/PhysRevA.87.033811}{Phys. Rev. A {\bf 87}, 033811}~(2013).

\bibitem{Ryl2}
S.~Ryl, J.~Sperling, and W.~Vogel.
\newblock ``Quantifying nonclassicality by characteristic functions''.
\newblock \href{https://dx.doi.org/10.1103/PhysRevA.95.053825}{Phys. Rev. A {\bf 95}, 053825}~(2017).

\bibitem{Bohmann}
Martin Bohmann, Elizabeth Agudelo, and Jan Sperling.
\newblock ``Probing nonclassicality with matrices of phase-space distributions''.
\newblock \href{https://dx.doi.org/10.22331/q-2020-10-15-343}{{Quantum} {\bf 4}, 343}~(2020).

\bibitem{horn2012matrix}
R.A. Horn and C.R. Johnson.
\newblock ``Matrix analysis''.
\newblock Cambridge University Press. ~(2012).

\bibitem{ManuilovMoslehianXu2020}
Vladimir Manuilov, Mohammad~Sal Moslehian, and Qingxiang Xu.
\newblock ``Douglas factorization theorem revisited''.
\newblock \href{https://dx.doi.org/10.1090/proc/14757}{Proceedings of the American Mathematical Society {\bf 148}, 1139--1151}~(2020).

\bibitem{vonNeumann1937}
John Von~Neumann.
\newblock ``Some matrix-inequalities and metrization of matrix-space''.
\newblock Tomsk Univ. Rev. {\bf 1}, 286--300~(1937).

\bibitem{CARLSSON2021149}
Marcus Carlsson.
\newblock ``von neumann’s trace inequality for hilbert–schmidt operators''.
\newblock \href{https://dx.doi.org/https://doi.org/10.1016/j.exmath.2020.05.001}{Expositiones Mathematicae {\bf 39}, 149--157}~(2021).

\bibitem{Hall2015}
Brian~C. Hall.
\newblock ``Lie groups, lie algebras, and representations: An elementary introduction''.
\newblock Volume 222 of Graduate Texts in Mathematics.
\newblock Springer. ~(2015).
\newblock 2 edition.

\bibitem{Schur}
J.~Schur.
\newblock ``Über potenzreihen, die im innern des einheitskreises beschränkt sind.''.
\newblock \href{https://dx.doi.org/doi:10.1515/crll.1917.147.205}{Journal für die reine und angewandte Mathematik {\bf 1917}, 205--232}~(1917).

\bibitem{MatrixCookbook}
K.~B. Petersen and M.~S. Pedersen.
\newblock ``The matrix cookbook''~(2012).
\newblock Version 2012.11.15.

\bibitem{Bittel2025optimalestimatesof}
Lennart Bittel, Francesco~Anna Mele, Antonio~Anna Mele, Salvatore Tirone, and Ludovico Lami.
\newblock ``Optimal estimates of trace distance between bosonic {G}aussian states and applications to learning''.
\newblock \href{https://dx.doi.org/10.22331/q-2025-06-12-1769}{{Quantum} {\bf 9}, 1769}~(2025).

\bibitem{UHLMANN1976273}
A.~Uhlmann.
\newblock ``The transition probability in the state space of a-algebra''.
\newblock \href{https://dx.doi.org/https://doi.org/10.1016/0034-4877(76)90060-4}{Reports on Mathematical Physics {\bf 9}, 273--279}~(1976).

\bibitem{Jozsa01121994}
Richard Jozsa.
\newblock ``Fidelity for mixed quantum states''.
\newblock \href{https://dx.doi.org/10.1080/09500349414552171}{Journal of Modern Optics {\bf 41}, 2315--2323}~(1994).

\bibitem{Banchi}
Leonardo Banchi, Samuel~L. Braunstein, and Stefano Pirandola.
\newblock ``Quantum fidelity for arbitrary gaussian states''.
\newblock \href{https://dx.doi.org/10.1103/PhysRevLett.115.260501}{Phys. Rev. Lett. {\bf 115}, 260501}~(2015).

\bibitem{Bures1969}
D.~Bures.
\newblock ``An extension of kakutani's theorem on infinite product measures to the tensor product of semifinite {$w^*$}-algebras''.
\newblock Transactions of the American Mathematical Society {\bf 135}, 199--212~(1969).

\bibitem{PhysRevA.77.062319}
Paulina Marian and Tudor~A. Marian.
\newblock ``Bures distance as a measure of entanglement for symmetric two-mode gaussian states''.
\newblock \href{https://dx.doi.org/10.1103/PhysRevA.77.062319}{Phys. Rev. A {\bf 77}, 062319}~(2008).

\bibitem{Holevo2024}
A.~S. Holevo.
\newblock ``On estimates of the bures distance between bosonic gaussian states''~(2024).
\newblock  \href{https://arxiv.org/abs/2412.04875}{arXiv:2412.04875}.

\bibitem{Petz1986}
D{\'e}nes Petz.
\newblock ``Quasi-entropies for finite quantum systems''.
\newblock Reports in Mathematical Physics {\bf 23}, 57--65~(1986).

\bibitem{MuellerLennert2013}
Martin M{\"u}ller-Lennert, Fr{\'e}d{\'e}ric Dupuis, Oleg Szehr, Serge Fehr, and Marco Tomamichel.
\newblock ``On quantum r{\'e}nyi entropies: a new definition and some properties''.
\newblock \href{https://doi.org/10.1063/1.4838856}{Journal of Mathematical Physics {\bf 54}, 122203}~(2013).

\bibitem{Wilde2014}
Mark~M. Wilde, Andreas Winter, and Dong Yang.
\newblock ``Strong converse for the classical capacity of entanglement-breaking and hadamard channels via a sandwiched r{\'e}nyi relative entropy''.
\newblock \href{https://doi.org/10.1007/s00220-014-2122-x}{Communications in Mathematical Physics {\bf 331}, 593--622}~(2014).

\end{thebibliography}

\appendix

\section{Wasserstein distance for non-centered Gaussian states}
\label{sec:noncentered}
Let us take two Gaussian states: $\rho_\gamma$  with covariance matrix $\gamma$ and mean  vector $\boldsymbol{\mu}$
and $\rho_\sigma$ with covariance matrix $\sigma$ and mean vector
and $\boldsymbol{\eta}$. The states  $\rho_\gamma^0$ and $\rho_\sigma^0$ are the Gaussian states with the same covariance matrices, but centered on the origin (so zero mean vector). Then recalling the definition of the covariance matrix in Eq.~\eqref{covmat}, we can compute the cost of a Gaussian coupling $\Pi$ between   $\rho_\gamma$  and  $\rho_\sigma$ (with non-zero displacement vector) as follows:
\eqarray{C(\Pi)&=&\tr[C\Pi]=\frac12\sum_i\tr[(\hat R_i\otimes\mathds{1}-\mathds{1}\otimes \hat R_i^\top)\Pi(\hat R_i\otimes\mathds{1}-\mathds{1}\otimes \hat R_i^\top)]\nonumber\\
&=&\frac12\sum_i\Big((\gamma_\Pi)_{i,i}+\mean{\hat R_{i}}^2- 2(\gamma_\Pi)_{i,n+i}-2\mean{\hat R_i}\mean{\hat R_{n+i}}+(\gamma_\Pi)_{n+i,n+i}+\mean{\hat R_{n+i}}^2\Big)\nonumber\\
&=&\frac12\tr[\gamma+\sigma-2X]+\frac12\sum_i\left(\mean{\hat R_i}-\mean{\hat R_{n+i}}\right)^2\nonumber\\
&=&\frac12\tr[\gamma+\sigma-2X]+\frac{1}{2}\sum_i\left({\boldsymbol{\mu}}_i-{\boldsymbol{\eta}}_i\right)^2\nonumber\\
&=&\frac12\tr[\gamma+\sigma-2X]+\frac12|{\boldsymbol{\mu}}-{\boldsymbol{\eta}}|^2
}
and the Wasserstein distance is given by
\begin{align}
    D^2(\rho_\gamma,\rho_\sigma)=\underset{\Pi\in\mathcal{C}(\rho_\gamma,\rho_\sigma)}{\text{min}}C(\Pi)=    D^2(\rho_\gamma^0,\rho_\sigma^0)+\frac12|{\boldsymbol{\mu}}-{\boldsymbol{\eta}}|^2.
\end{align}
Moreover, if $\Pi_{\mathrm{opt}}$ denotes the quantum optimal transport plan between $\rho_\gamma$ and $\rho_\sigma$, it has the covariance matrix $\gamma_\Pi^{opt}$ and the mean vector $(\boldsymbol{\mu},\boldsymbol{\eta})$. Then $\Pi_{\mathrm{opt}}^0$ denotes the quantum optimal transport plan between their centered counterparts $\rho_\gamma^0$ and $\rho_\sigma^0$; it has the same covariance matrix and a zero-mean vector.

\section{Shape of $X$}
\label{Appendix Shape of X}

In \cite{DePalma2024}, it was argued that the optimal matrix $X$ in the covariance matrix \eqref{CouplingCovarianceMatrix} was necessarily diagonal and could always be taken as proportional to the identity, that is $X'=\frac12\tr[X]\mathds{1}$, but no details were given. We demonstrate here why this holds, but emphasize that this is only true for covariance matrices proportional to the identity, i.e., thermal states. 

Defining $X'=\frac12\tr[X]\mathds{1}$ does not change the cost, so the question is simply: is the new coupling still physical (so we have a valid solution)? For this to be true, we need to show that 
\eq{\begin{pmatrix}
    A & X'\\ X'^\top&B
\end{pmatrix}\geq\pm\frac{i}{2}\begin{pmatrix}
    \Omega&0\\0&-\Omega
\end{pmatrix}\label{URXprime}}
meaning it still satisfies the uncertainty principle.
According to the singular value decomposition, $X$ can be written as $\mathcal U D\mathcal V^\top$ where $\mathcal  U$ and $\mathcal  V$ are orthogonal matrices ($\mathcal U^\top\mathcal U=\mathcal V^\top\mathcal V=\mathds{1}$)  and $D$ is diagonal (with the singular values on the diagonal). We can thus define the symplectic transformation
\begin{align}
\begin{pmatrix}
    \mathcal U&0\\0&\mathcal V
\end{pmatrix}\quad\text{with}\quad
\begin{pmatrix}
    \mathcal U&0\\0&\mathcal V
\end{pmatrix} \begin{pmatrix}
    \Omega&0\\0&-\Omega
\end{pmatrix} \begin{pmatrix}
    \mathcal U^\top&0\\0&\mathcal V^\top
\end{pmatrix}=\begin{pmatrix}
    \Omega&0\\0&-\Omega
\end{pmatrix}\ \ \ \text{and}\quad\det \begin{pmatrix}
    \mathcal U&0\\0&\mathcal V
\end{pmatrix}=1,
\end{align}
such that 
\begin{align}
    \begin{pmatrix}
   \mathcal  U^\top&0\\0&\mathcal V^\top
\end{pmatrix}
\begin{pmatrix}
    A & X\\ X^\top&B
\end{pmatrix}
\begin{pmatrix}
   \mathcal  U&0\\0&\mathcal V
\end{pmatrix}= \begin{pmatrix}
   \mathcal  U^\top A \mathcal U & D\\ D^\top&\mathcal V^\top B\mathcal V
\end{pmatrix}
\underset{ A=\nu_1\mathds{1},\ B=\nu_2\mathds{1}}{=}
\begin{pmatrix}
    A & D\\ D^\top&B
\end{pmatrix},
\end{align}
where the last equality only holds because covariance matrices $A$ and $B$ are proportional to the identity. 
Now, if $\gamma_\Pi$ is physical, then
\eqarray{&
\begin{pmatrix}
    A & X\\ X^\top&B
\end{pmatrix}\geq \frac{i}{2}\begin{pmatrix}
    \Omega&0\\0&-\Omega
\end{pmatrix}\nonumber\\
\Leftrightarrow &
\begin{pmatrix}
    \mathcal U^\top&0\\0&\mathcal V^\top
\end{pmatrix}
\begin{pmatrix}
    A & X\\ X^\top&B
\end{pmatrix}
\begin{pmatrix}
  \mathcal U&0\\0&\mathcal V
\end{pmatrix}\geq\frac{i}{2}\begin{pmatrix}
    \mathcal U^\top&0\\0&\mathcal V^\top
\end{pmatrix} \begin{pmatrix}
    \Omega&0\\0&-\Omega
\end{pmatrix} \begin{pmatrix}
    \mathcal U&0\\0&\mathcal V
\end{pmatrix}\nonumber\\
\Leftrightarrow & \begin{pmatrix}
    A & D\\ D^\top&B
\end{pmatrix}\geq \frac{i}{2}\begin{pmatrix}
    \Omega&0\\0&-\Omega
\end{pmatrix}.
}
So $X$ can be assumed to be diagonal when $A$ and $B$ are proportional to identity because the coupling is still a physical coupling. 
But it is important to mention that for general Gaussian states, even squeezed thermal states, the argument does not hold anymore, and one cannot assume $X$ to be diagonal.

\section{Some lemmas leading to the computation of the Wasserstein distance for Gaussian states}
\label{Appendix:Lemmas}
This section contains different Lemmas that are needed to compute the Wasserstein distance between two Gaussian states.

\begin{mdframed}[backgroundcolor=quantumviolet!5!,linecolor=quantumviolet!5!]
\begin{lemma}\label{lemma:defMatrices}
    \textbf{Some matrix definitions} 
    \cite{horn2012matrix}\vspace{-10pt}
\begin{itemize}
    \item A matrix is \textbf{symmetric} if $M=M^\top$.
    \item A matrix is \textbf{skew-symmetric} if $M=-M^\top$.
     \item A matrix is \textbf{hermitian} if $M=M^\dag$.
     \item Every hermitian matrix $M$ can be written as $M=S+iK$ where $S=\frac12(M+M^*)$ is a symmetric matrix and $K=\frac1{2i}(M-M^*)$ is skew-symmetric.
     \item    The \textbf{trace-class (nuclear) norm} of a matrix,  
$\|M\|_{1} = \operatorname{Tr}\left(\sqrt{M^\dagger M}\right)$
is equal to the sum of its singular values, denoted $\sigma_j(M)$. If $M$ is positive semi-definite, $\|M\|_1=~\tr (M)$.
\item  The \textbf{operator norm}  of a matrix, $\|M\|_{op}$  is equal to the greatest singular value of $M$. 
\end{itemize}
\end{lemma}
\end{mdframed}

\begin{mdframed}[backgroundcolor=quantumviolet!5!,linecolor=quantumviolet!5!]
\begin{lemma}\label{lemma:Douglas} 
\textbf{The Douglas Factorization} \cite{ManuilovMoslehianXu2020}

For two square matrices $A$ and $B$ we have
$$A^\dag A\geq B^\dag B\qquad\Leftrightarrow\qquad B=FA$$ where $F$ is a contraction, meaning that $F^\dag F\leq I$.
\end{lemma}
\end{mdframed}

\begin{mdframed}[backgroundcolor=quantumviolet!5!,linecolor=quantumviolet!5!]
\begin{lemma} \label{lemma:Mpositive} $ $

Let $M=S+iK$ be a hermitian matrix with $S$ and $K$ assumed to be invertible. Then, $M$ is positive definite ($M>0$) if and only if
\vspace{-10pt}
\begin{enumerate}
    \item $S> 0$,
    \item $-\left(\sqrt{S}^{-1}K\sqrt{S}^{-1}\right)^2<I
       \iff
\left|\sqrt{S}^{-1}K\sqrt{S}^{-1}\right|> I
       \iff
    S+KS^{-1}K>0
         \iff
            K^{-1}SK^{-1}+S^{-1}< 0.$
        % \item One can write $K=\sqrt{S}F\sqrt{S}$ with $F$ a skew symmetric matrix such that $-F^2<I$.
\end{enumerate}
\end{lemma}
\end{mdframed}

\begin{proof}
 Note first that $S$ is symmetric and $K$ is skew-symmetric.
    \begin{enumerate}
        \item Since $M>0$, we also have $M^*>0$. Therefore, $S=\frac12(M+M^*)>0$.
        \item Since $S>0$, $S^{-1}$ exist and $\sqrt{S}>0$. Then, we can write
\eq{M=S+iK=\sqrt{S}\left(I+i\sqrt{S}^{-1}K\sqrt{S}^{-1}\right)\sqrt{S}>0\quad\iff\quad I+i\sqrt{S}^{-1}K\sqrt{S}^{-1}>0.}
Therefore, we also have $I-i\sqrt{S}^{-1}K\sqrt{S}^{-1}>0$ and 
\eq{0<\left(I+i\sqrt{S}^{-1}K\sqrt{S}^{-1}\right)\left(I-i\sqrt{S}^{-1}K\sqrt{S}^{-1}\right)=I+\left(\sqrt{S}^{-1}K\sqrt{S}^{-1}\right)^2.}
That is the first inequality. Since $K$ is skew-symmetric and $S$ is symmetric, $\sqrt{S}^{-1}K\sqrt{S}^{-1}$ is skew-symmetric.
That yields to 
\begin{align}
    I&>-\left(\sqrt{S}^{-1}K\sqrt{S}^{-1}\right)^2\!=\left(\sqrt{S}^{-1}K\sqrt{S}^{-1}\right)^\top\!\left(\sqrt{S}^{-1}K\sqrt{S}^{-1}\right)=\left|\sqrt{S}^{-1}K\sqrt{S}^{-1}\right|^2\nonumber\\
    \Leftrightarrow\quad0&<I-\left|\sqrt{S}^{-1}K\sqrt{S}^{-1}\right|^2=\left(-\left|\sqrt{S}^{-1}K\sqrt{S}^{-1}\right|\right)\left(I+\left|\sqrt{S}^{-1}K\sqrt{S}^{-1}\right|\right)\nonumber\\
    \Leftrightarrow\quad0&<I-\left|\sqrt{S}^{-1}K\sqrt{S}^{-1}\right|<I+\left|\sqrt{S}^{-1}K\sqrt{S}^{-1}\right|
\end{align}
where the last line comes from the fact that, by definition, the norm of a matrix is positive. That is the second inequality.
The other equivalent inequalities are obtained as follows:
\eq{I>-\left(\sqrt{S}^{-1}K\sqrt{S}^{-1}\right)^2=-\sqrt{S}^{-1}KS^{-1}K\sqrt{S}^{-1}} also implies $S>-KS^{-1}K$ and $K^{-1}SK^{-1}>-S^{-1}$.
    \end{enumerate}
\end{proof}
NOTE: The above lemma will later be applied to covariance matrices. We restrict our attention to strictly positive matrices, since equality in the uncertainty principle is attained only by pure Gaussian states. Because our analysis focuses on mixed states (the pure case being treated separately), the strictly positive regime is the relevant one.

\begin{mdframed}[backgroundcolor=quantumviolet!5!,linecolor=quantumviolet!5!]
\begin{lemma}\label{VNineq2}
\textbf{The Von Neumann inequality} \cite{vonNeumann1937,CARLSSON2021149}

The absolute value of the trace of the product of two matrices is less than or equal to the sum of the products of their singular
values, arranged in descending order:
$$|\tr(XY)|\leq\sum_j\sigma_j(X)\sigma_j(Y)\leq\|X\|_{op}\sum_j\sigma_j(Y)=\|X\|_{op}\| Y\|_1.$$
In addition, the first inequality is saturated if and only if $X$ and $Y^\dag$ share singular vectors, i.e. they are aligned.
\end{lemma}
\end{mdframed}

\begin{mdframed}[backgroundcolor=quantumviolet!5!,linecolor=quantumviolet!5!]
\begin{lemma}\label{Lemma:optimizationOverPostiveW}$ $

    Let $W$ be a real matrix such that $W^\top W\leq\mathds 1$ and 
   $$ W^\top\left(\mathds{1}\pm  \frac{i}{b} \Omega \right) W \leq\mathds{1}\pm  \frac{i}{a}\Omega$$
   then, the problem of maximizing $\tr[W]$ can be reduced to matrices $W$ that are  positive semi-definite symmetric matrices.
\end{lemma}
\end{mdframed}
\begin{proof}  
    Let us choose a matrix $\tilde{W}$ in the feasible set (i.e. it satisfies both inequalities) and let us apply a polar decomposition $\tilde{W}=\mathcal U P$ where $\mathcal U$ is orthogonal (because $\tilde{W}$ is real) and $P$ is positive semi-definite and symmetric \cite{Hall2015}. Then, we have $\mathds 1\geq\tilde{W}^\top \tilde W=P^\top \mathcal U^\top \mathcal U P=P^\top P$ and
\begin{align}
  &  W^\top\left(\mathds{1}\pm  \frac{i}{b} \Omega \right) W \leq\mathds{1}\pm  \frac{i}{a}\Omega\nonumber\\
  \Leftrightarrow\qquad  &  P^\top \mathcal U^\top\left(\mathds{1}\pm  \frac{i}{b} \Omega \right) \mathcal UP \leq\mathds{1}\pm  \frac{i}{a}\Omega\nonumber\\
    \Leftrightarrow\qquad   &  P^\top \left(\mathds{1}\pm  \frac{i\det \mathcal U}{b} \Omega \right) P \leq\mathds{1}\pm  \frac{i}{a}\Omega\nonumber\\
        \Leftrightarrow\qquad   &  P^\top \left(\mathds{1}\pm  \frac{i}{b} \Omega \right) P \leq\mathds{1}\pm  \frac{i}{a}\Omega,
\end{align}
since $\det \mathcal U=\pm 1$. Therefore, $P$ is also in the feasible set. Finally, we have
\eq{\tr[\tilde W]=\tr[\mathcal UP]\leq|\tr[\mathcal UP]|\leq\tr[P]}
where we use the Von Neumann inequality in the last line (Lemma.~\eqref{VNineq2}). Hence, to maximize the trace, we can restrict the search to symmetric positive semi-definite matrices.
\end{proof}

\begin{mdframed}[backgroundcolor=quantumviolet!5!,linecolor=quantumviolet!5!]
\begin{lemma}
\label{COR-SMALL1}$ $

Let $K$ be a real symmetric matrix and 
$\eta>\frac{|\zeta|}{2}>0 $. 
Then the problem
$$
 \max \tr[K]=\max(a + d) \qquad
\text{such that  }\qquad 
0\leq  
K=
\begin{pmatrix}
a & b \\ 
b & d
\end{pmatrix}
\le \eta \mathds{1} + \operatorname{i}\frac{\zeta}{2}\Omega,
$$
 has  the unique solution $K^*=(\eta - \tfrac{|\zeta|}{2})\mathds{1}$.
\end{lemma}
\end{mdframed}
\begin{proof}
Let us start with $\eta=1$. Since $K\geq0$, we have $b^2\leq ad$. The other side of the constraint $K\leq \mathds{1}+i\frac{\zeta}{2}\Omega$ can be read as
\begin{align}
\begin{pmatrix}
1 - a & \operatorname{i}\tfrac{\zeta}{2} - b\\
-\operatorname{i}\tfrac{\zeta}{2} - b & 1 - d
\end{pmatrix} 
\ge 0
\end{align}
which implies $0 \le a,d \le 1$ and $(1-a)(1-d) \ge b^2 + \frac{\zeta^2}{4}.$
Thus, the feasible set is
\eq{
F_b = \left\{(a,d)\quad :\quad 0 \le a,d \le 1, \quad b^2 \le ad, \quad (1-a)(1-d) \ge b^2 + \tfrac{\zeta^2}{4}\right\}.
}
For $b_i$'s such that $|b_1|> |b_2| \geq  0$, the feasible sets satisfy $F_{b_1} \subseteq F_{b_2}$.  
Hence, $b$ with the smallest $|b|$ gives the largest feasible set.  
We can therefore set $b = 0$ and write
\eq{
F_0 = \left\{(a,d)\quad :\quad 0 \le a,d \le 1,  \quad (1-a)(1-d) \ge \tfrac{\zeta^2}{4}\right\}.
}
We thus want to solve
\eq{
\max (a + d) = \max_{(a,d) \in F_0} (a+d).
}
Defining 
$  a_0=1-a,\  d_0=1-d$, the problem 
becomes
\eq{
 \max_{(a,d) \in F_0} (a+d)= 2-\min_{(a_0,d_0) \in G_0} (a_0+d_0)\quad\text{with}\quad G_0 = \left\{(a_0,d_0)\quad :\quad 0 \le a_0,d_0 \le 1,  \quad a_0d_0 \ge \tfrac{\zeta^2}{4}\right\}.
}
The minimum  happens at $a_0 = d_0 = \frac{|\zeta|}{2}$,
so the maximum of the inital problem occurs at $a = d = 1 - \frac{|\zeta|}{2}$,
giving $
\max(a + d) = 2 - |\zeta|$ and 
\eq{
K^\star =
\begin{pmatrix}
a & b \\ b & d
\end{pmatrix}
= (1 - \tfrac{|\zeta|}{2})I\label{solutionK}.
}
We can now scale the problem and consider values of $\eta$ different from 1.
We define
\eq{
 \tilde{K}=\frac{1}{\eta}K =
\begin{pmatrix}
 \tilde{a} &  \tilde{b}\\
 \tilde{b} &  \tilde{d}
\end{pmatrix},
\quad \text{where }\quad  \tilde{a} = \frac{a}{\eta}, \quad  \tilde{d} = \frac{d}{\eta}, \quad  \tilde{b} = \frac{b}{\eta}
}
and the problem now reads
\eq{
 \max (a + d)=\eta\max(\tilde a+\tilde d) \qquad
\text{such that } \qquad
0\leq  
\tilde K=
\le  \mathds{1} + \operatorname{i}\frac{\tilde \zeta}{2}\Omega, \quad \text{where }\tilde{\zeta} = \frac{\zeta}{\eta}.
}
According to Eq.~\eqref{solutionK}, the solution is given by
 \eq{
\tilde K^\star
= (1 - \tfrac{|\tilde \zeta|}{2})\mathds{1}=(1 - \tfrac{ |\zeta|}{2\eta})\mathds{1}\qquad\Leftrightarrow\qquad K^*=\eta K=(\eta - \tfrac{ |\zeta|}{2})\mathds{1}.
}
\end{proof}

\begin{mdframed}[backgroundcolor=quantumviolet!5!,linecolor=quantumviolet!5!]
\begin{lemma}\label{LEM-2b2-GEN-FORM} $ $

Let $K$ be a real symmetric matrix. Then, the problem
$$
 \max\tr [K]
 %=\max (a + d) 
 \qquad
\text{such that } \qquad
0\leq  
K
\qquad\text{and}\quad  
 K\left( \mathds{1} -\operatorname{i}\frac{a_{\mathtt{0}}}{2}\Omega\right)K\leq    \mathds{1} -\operatorname{i}\frac{b_{\mathtt{0}}}{2} \Omega
$$
with $|a_0|,|b_0|\leq2$ has the unique solution 
\begin{align*}
    K^\star=\left\{\begin{array}{ccc}
          \frac{\sqrt{2+b_0}}{\sqrt{2+a_0}}\mathds{1} & \text{if }& a_0\geq b_0 \\
          & &\\
     \frac{\sqrt{2-b_0}}{\sqrt{2-a_0}}\mathds{1} & \text{if }& a_0\leq b_0.
    \end{array}\right.
\end{align*}
\end{lemma}
\end{mdframed}

\begin{proof}
We start by multiplying both sides of the constraint by $\sqrt{\mathds{1}-\operatorname{i}\frac{a_{\mathtt{0}}}{2}\Omega}$ and obtain:
\begin{align}
    \sqrt{\mathds{1}-\operatorname{i}\frac{a_{\mathtt{0}}}{2}\Omega}\, K\left( \mathds{1}-\operatorname{i}\frac{a_{\mathtt{0}}}{2}\Omega\right)K\sqrt{\mathds{1}-\operatorname{i}\frac{a_{\mathtt{0}}}{2}\Omega}&\leq \sqrt{\mathds{1}-\operatorname{i}\frac{a_{\mathtt{0}}}{2}\Omega} \left(  \mathds{1}-\operatorname{i}\frac{b_{\mathtt{0}}}{2} \Omega\right)\sqrt{\mathds{1}-\operatorname{i}\frac{a_{\mathtt{0}}}{2}\Omega}\nonumber\\
   \left( \sqrt{\mathds{1}-\operatorname{i}\frac{a_{\mathtt{0}}}{2}\Omega}\, K\,  \sqrt{\mathds{1}-\operatorname{i}\frac{a_{\mathtt{0}}}{2}\Omega}\right)^2&\leq \sqrt{\mathds{1}-\operatorname{i}\frac{a_{\mathtt{0}}}{2}\Omega} \left(  \mathds{1}-\operatorname{i}\frac{b_{\mathtt{0}}}{2} \Omega\right)\sqrt{\mathds{1}-\operatorname{i}\frac{a_{\mathtt{0}}}{2}\Omega}.
\end{align}
Since $K\geq0$, $\sqrt{\mathds{1}-\operatorname{i}\frac{a_{\mathtt{0}}}{2}\Omega}\, K\,  \sqrt{\mathds{1}-\operatorname{i}\frac{a_{\mathtt{0}}}{2}\Omega}\geq0$ as long as $|a_0|\leq2$. 
Then, taking the square root, multiplying again both sides by $\sqrt{\mathds{1}-\operatorname{i}\frac{a_{\mathtt{0}}}{2}\Omega}^{-1}$, and using the fact that matrices of the form $\mathds{1}-i\frac{x}{2}\Omega$ commute, we  get
\begin{align}
K&\leq\sqrt{\mathds{1}-\operatorname{i}\frac{a_{\mathtt{0}}}{2}\Omega}^{-1}\sqrt{ \sqrt{\mathds{1}-\operatorname{i}\frac{a_{\mathtt{0}}}{2}\Omega} \left(  \mathds{1}-\operatorname{i}\frac{b_{\mathtt{0}}}{2} \Omega\right)\sqrt{\mathds{1}-\operatorname{i}\frac{a_{\mathtt{0}}}{2}\Omega}}\sqrt{\mathds{1}-\operatorname{i}\frac{a_{\mathtt{0}}}{2}\Omega}^{-1}\nonumber\\
&=\sqrt{\left(\mathds{1}-\operatorname{i}\frac{a_{\mathtt{0}}}{2}\Omega\right)^{-1}\left(\mathds{1}-\operatorname{i}\frac{b_{\mathtt{0}}}{2} \Omega\right) }.
\end{align}
We can compute the inverse of $(\mathds{1}-\operatorname{i}\frac{a_{\mathtt{0}}}{2}   \Omega)$ by defining $(\mathds{1}-\operatorname{i}\frac{a_{\mathtt{0}}}{2}   \Omega)^{-1}=R+iQ$. This implies
\begin{align}
    \left(\mathds{1}-i\frac{a_0}{2}\Omega\right)(R+iQ)=\mathds{1}&\quad\iff\quad
    %R+\frac{a_0}{2}\Omega Q+i\left(Q-\frac{a_0}{2}\Omega R\right)&=\mathds{1}\\
    R+\frac{a_0}{2}\Omega Q =\mathds 1 \ \ \text{and}\ \ Q-\frac{a_0}{2}\Omega R=0\nonumber\\
    &\quad\iff\quad   Q=\frac{a_0}{2}\Omega R\ \  \text{and}\ \ R =\mathds 1 +\frac{a_0^2}{4} R \nonumber\\
   &\quad\iff\quad    R=\frac1{1-\frac{a_0^2}{4}}\mathds 1\ \ \text{and} \ \ Q=\frac1{1-\frac{a_0^2}{4}}\frac{a_0}{2}\Omega
\end{align}
which gives
$ \left(\mathds{1}-\operatorname{i}\frac{a_{\mathtt{0}}}{2}   \Omega\right)^{-1}=\left(1 - \frac{a_{\mathtt{0}}^2}{4}   \right)^{-1}\left(\mathds{1}+\operatorname{i}\frac{a_{\mathtt{0}}}{2}    \Omega\right).$
Hence, the constraint on the matrix $K$ reads
\begin{align}
   K&\leq \frac{1}{\sqrt{1 - \frac{a_{\mathtt{0}}^2}{4}}} \sqrt{\left(\mathds{1}+\operatorname{i}\frac{a_{\mathtt{0}}}{2} \Omega\right)\left(\mathds{1}-\operatorname{i}\frac{b_{\mathtt{0}}}{2}  \Omega\right) } = \frac{1}{\sqrt{1 \!-\! \frac{a_{\mathtt{0}}^2}{4}}} \sqrt{\left(1-\frac{a_0b_0}{4}\right)\!\mathds{1}+\operatorname{i}\frac{a_0-b_0}{2} \Omega }\\
  &=\frac{1}{4\sqrt{1\! -\! \frac{a_{\mathtt{0}}^2}{4}}}\left[\!\left(\!\sqrt{(2\!+\!a_0)(2\!-\!b_0)}\!+\!\sqrt{(2\!-\!a_0)(2\!+\!b_0)}\right)\mathds{1}
%  \right.\\ &\qquad\qquad\qquad\left.
  \!+\!\left(\sqrt{(2\!+\!a_0)(2\!-\!b_0)}\!-\!\sqrt{(2\!-\!a_0)(2\!+\!b_0)}\right)i \Omega
  \right]\!.\nonumber
\end{align}
We can now apply Lemma \ref{COR-SMALL1} to find the optimal $K^*$ with 
\begin{align}
    \eta&=\sqrt{(a_0+2)(-b_0+2)}+\sqrt{(-a_0+2)(b_0+2)}\nonumber\\
    \frac{\zeta}{2}&=\sqrt{(a_0+2)(-b_0+2)}-\sqrt{(-a_0+2)(b_0+2)}\ \left\{\begin{array}{ccc}
          \geq0  & \text{if }& a_0\geq b_0 \\
        \leq 0  & \text{if }& a_0\leq b_0.
    \end{array}\right.
\end{align}
Note that $\eta\geq|\zeta|/2$ is always verified. 
Then, 
\begin{align}
    K^\star&=\frac{1}{4\sqrt{1 \!-\! \frac{a_{\mathtt{0}}^2}{4}}}\left[\!\left(\!\sqrt{(2\!+\!a_0)(2\!-\!b_0)}\!+\!\sqrt{(2\!-\!a_0)(2\!+\!b_0)}\right)
    %\right.\\ &\qquad\qquad\qquad\left.
    \!-\!\left|\sqrt{(2\!+\!a_0)(2\!-\!b_0)}\!-\!\sqrt{(2\!-\!a_0)(2\!+\!b_0)}\right|\right]\mathds{1}\nonumber
    \\
    &=\left\{\begin{array}{ccc}
          \frac{1}{2\sqrt{1 -\frac{a_{\mathtt{0}}^2}{4} }}\left( \sqrt{(2-a_0)(2+b_0)}\right)\mathds{1}  & \text{if }& a_0\geq b_0 \\
     \frac{1}{2\sqrt{1 -\frac{a_{\mathtt{0}}^2}{4} }}\left( \sqrt{(2+a_0)(2-b_0)}\right)\mathds{1}  & \text{if }& a_0\leq b_0
    \end{array}\right.\nonumber\\
        &=\left\{\begin{array}{ccc}
          \frac{\sqrt{2+b_0}}{\sqrt{2+a_0}}\mathds{1} & \text{if }& a_0\geq b_0 \\
     \frac{\sqrt{2-b_0}}{\sqrt{2-a_0}}\mathds{1} & \text{if }& a_0\leq b_0.
    \end{array}\right.
\end{align}
\end{proof}

\begin{mdframed}[backgroundcolor=quantumviolet!5!,linecolor=quantumviolet!5!]
\begin{lemma}\label{LEM-SOL-SOL}$ $

For the set \begin{align*}
    \mathcal{S}=\left\{X\ :\ X^\top A^{-1} X\leq B\qquad\text{and} \qquad X \text{ symmetric, positive semi-definite}\right\}
\end{align*}
where $A$ and $B$ are positive definite matrices, we consider the two following optimization problems.  Let
    \begin{align*}
(P1)&  \quad  \max\left\{\operatorname{Tr}(X{N})\ :\ X\in\mathcal{S}\right\}\qquad\text{and}\qquad (P2)  \quad \max\left\{\operatorname{Tr}(X)\ :\ X\in\mathcal{S}\right\}
\end{align*}
where $N$ is also positive definite. Then, 
if   $X^\star=x^\star \mathds{1}$ maximizes  $(P2)$, $X^\star$ is also a maximizer of $(P1)$, and the explicit  solution to $(P1)$ is given by 
    \begin{align*}
 \max\left\{\operatorname{Tr}(X{N})\ :\ X\in\mathcal{S}\right\}=\operatorname{Tr}(x^\star N).
\end{align*}
\end{lemma}
\end{mdframed}
\begin{proof}
According to Lemma \ref{VNineq2}, we have 
\begin{align}    
\max \tr(XN)&\leq\max|\tr(XN)|\leq\max\left(\sum_j\sigma_j(X)\sigma_j(N)\right)=\max\left(\sum_j\lambda_j(X)\lambda_j(N)\right)
\end{align}
the last inequality being true because $X$ and $N$  are semi-definite positive.
The maximum value of the sum $\sum_{j} \lambda_j(X)\lambda_j(N)$ is bounded by the available "size" of $X$ (its nuclear norm) and the weighting $N$. Since $X^\star$ is the maximizer of the unweighted trace, $\operatorname{Tr}(X^\star) \geq \operatorname{Tr}(X)$, and $X^\star$ is proportional to the identity ($\lambda_j(X^\star) = x^\star$ for all $j$), the maximum is achieved exactly at $X^\star$. 
 In other words, the maximum is attained when the matrix $X$ both maximizes its total size ($\operatorname{Tr}(X)$) and is optimally aligned (share the same eigenvectors) with $N$ (again according to Lemma \ref{VNineq2}). Now, $X^*$ is proportional to identity so commutes with every matrix, including $N$, which ensures it satisfies the necessary alignment conditions for maximizing  $\operatorname{Tr}(XN)$.

 Any other feasible $X'$ must satisfy $\sum_{j} \lambda_j(X') \leq 2m x^\star$. To achieve 
a higher value of $\sum_{j} \lambda_j(X')\lambda_j(N)$, $X'$ would need to trade its smaller eigenvalues for larger ones to align with the larger $\lambda_j(N)$, while staying inside the feasible set $\mathcal{S}$ (remember that in the inequality above, the eigenvalues are arranged in descending order). Since the feasible set $\mathcal{S}$ is centered around $X=0$, and $X^\star$ is already the largest sphere inside it, the constraints penalize any attempt to create a larger eigenvalue so the sum $\sum_{j} \lambda_j(X')\lambda_j(N)$ cannot exceed $x^\star \operatorname{Tr}(N)$.
Therefore, 
\begin{align*} 
\max \tr(XN)\leq\max\left(\sum_j\lambda_j(X)\lambda_j(N)\right)
=x^*\sum_j\lambda_j(N)=x^*\tr(N)=\tr(X^*N)
\end{align*}
which implies that $X^*$ is the maximizer of (P1).
\end{proof}

\section{Physicality of the optimal coupling}
\label{Appendix:PhysicalityOptimalX}
Because covariance have to verify the uncertainty principle, Eq.~\eqref{CouplingPhysical} imposes some  restrictions on $X$. To better understand them, we can describe $\gamma_\Pi^\pm$ using the Schur complement \cite{Schur,MatrixCookbook}. We get
\begin{align}\label{SCHUR}
   \gamma_\Pi^\pm =\gamma_\Pi\pm\frac{i}{2}\begin{pmatrix}
\Omega    &   0\\
 &\\
  0   &  -\Omega \\\end{pmatrix}=\begin{pmatrix}
 A \pm\frac{i}{2}\Omega    &   X\\
 &\\
  X^\top    &   B\mp\frac{i}{2}\Omega \\\end{pmatrix}={G^\pm }^\top \begin{pmatrix}
 \mathds{1} &  0\\
 &\\
 0   &   S^\pm \\
    \end{pmatrix}  G^\pm
\end{align}
where the Schur complement is defined as \footnote{Note that $A^\pm$ has to be invertible so it cannot be used for a pure state where we would have $A^\pm=0$.}
\begin{align}
S^\pm =B\mp\frac{i}{2}\Omega -X^\top\left(A\pm\frac{i}{2}\Omega \right)^{-1} X
\qquad\text{and}\qquad
G^\pm =
\begin{pmatrix}
\sqrt{A^\pm }   &     {\sqrt{A^\pm }}^{-1}X \\
 &\\
0 & \mathds{1}\\
    \end{pmatrix}.
\end{align}
A property of the Schur complement is that $\gamma_\Pi^\pm$ is positive definite if and only if $A^\pm$ and its complement $S^\pm$ are both positive definite. By definition $A^\pm\geq0$ so it remains to make sure that $S^\pm\geq0$ which implies $B^\mp\geq X^\top{A^\pm }^{-1} X$.
Using the optimal $X$ in Eq.~\eqref{solutionOptimlaX}, it means we want $\left(B\mp\frac{i}{2}\Omega\right)-X^{*\top}\left(A\pm\frac{i}{2}\Omega\right)^{-1}X^*\geq0$.
We can compute
\begin{align}
 &  X^{*\top}\left(A\pm\frac{i}{2}\Omega\right)^{-1}X^*=
     X^{*\top}\frac{1}{\nu_A^2-\frac14}\left(\nu_A^2A^{-1}\mp\frac{i}{2}\Omega\right)X^*&\\
  & =   \left\{\!\!\begin{array}{ccc}
\frac{(2\nu_B-1)(2\nu_A+1)}{4\nu_A\nu_B\left(\nu_A^2-\frac14\right)}
\sqrt{B}\sqrt{\!\sqrt{\!B}A\sqrt{\!B}} \sqrt{\!B^{-1}}\left(\nu_A^2A^{-1}\mp\frac{i}{2}\Omega\right)\sqrt{B^{-1}}\sqrt{\!\sqrt{\!B}A\sqrt{\!B}}\sqrt{\!B}& \text{if }& \nu_A\geq \nu_B \\
          &&\\
\!\frac{(2\nu_B+1)(2\nu_A-1)}{4\nu_A\nu_B\left(\nu_A^2-\frac14\right)}
\sqrt{B}\sqrt{\!\sqrt{\!B}A\sqrt{\!B}} \sqrt{\!B^{-1}}\left(\nu_A^2A^{-1}\mp\frac{i}{2}\Omega\right)\sqrt{B^{-1}}\sqrt{\!\sqrt{\!B}A\sqrt{\!B}}\sqrt{\!B} & \text{if }& \nu_A\leq \nu_B.
    \end{array}\right.&\nonumber
\end{align}
Noting that $\sqrt{B}\sqrt{\sqrt{B}A\sqrt{B}} \sqrt{B^{-1}}A^{-1}\sqrt{B^{-1}}\sqrt{\sqrt{B}A\sqrt{B}}\sqrt{B}=\sqrt{B}N N^{-1}N^{-1}N\sqrt{B}=B$ and using again the fact that $M\Omega M^\top= (\det M) \Omega$ for $2\times 2$ matrices, the inequality simplifies as
\begin{align}
    &\qquad\ \  \left\{\begin{array}{ccc}
 \left(B\mp\frac{i}{2}\Omega\right)- \frac{(2\nu_B-1)(2\nu_A+1)}{4\nu_A\nu_B\left(\nu_A^2-\frac14\right)}
\left(\nu_A^2 B\mp\frac{i}{2}\nu_A\nu_B\Omega\right)\geq0& \text{if }& \nu_A\geq \nu_B \\
          &&\\
\left(B\mp\frac{i}{2}\Omega\right)-\frac{(2\nu_B+1)(2\nu_A-1)}{4\nu_A\nu_B\left(\nu_A^2-\frac14\right)}
\left(\nu_A^2B\mp\frac{i}{2}\nu_A\nu_B\Omega\right)\geq0& \text{if }& \nu_A\leq \nu_B
    \end{array}\right.&\nonumber\\
 & \iff    \left\{\begin{array}{ccc}
 \left(\mathds1\mp\frac{i}{2\nu_B}\Omega\right)- \frac{(2\nu_B-1)(2\nu_A+1)}{4\nu_A\nu_B\left(\nu_A^2-\frac14\right)}
\left(\nu_A^2 \mathds1\mp\frac{i}{2}\nu_A\Omega\right)\geq0& \text{if }& \nu_A\geq \nu_B \\
          &&\\
\left(\mathds1\mp\frac{i}{2\nu_B}\Omega\right)-\frac{(2\nu_B+1)(2\nu_A-1)}{4\nu_A\nu_B\left(\nu_A^2-\frac14\right)}
\left(\nu_A^2\mathds1\mp\frac{i}{2}\nu_A\Omega\right)\geq0& \text{if }& \nu_A\leq \nu_B
    \end{array}\right.&\nonumber\\
 &  \iff      \left\{\begin{array}{ccc}
\left(1-\frac{(\nu_B-\frac12)}{\nu_B\left(\nu_A-\frac12\right)}\nu_A\right)\mathds1\mp\frac{i}{2\nu_B}\Omega\left(\frac{\nu_A-\nu_B}{\nu_A-\frac12}\right)
\geq0& \text{if }& \nu_A\geq \nu_B \\
          &&\\
\left(1-\frac{(\nu_B+\frac12)}{\nu_B\left(\nu_A+\frac12\right)}\nu_A\right)\mathds1\mp\frac{i}{2\nu_B}\Omega\left(\frac{\nu_A-\nu_B}{\nu_A+\frac12}
\right)\geq0& \text{if }& \nu_A\leq \nu_B
    \end{array}\right.&\nonumber\\
  &  \iff     \left\{\begin{array}{ccc}
\nu_B\left(\frac{\nu_A-\frac12}{\nu_A-\nu_B}\right)
\left(\frac{\nu_B\left(\nu_A-\frac12\right)-\nu_B\nu_A+\frac12\nu_A}{\nu_B\left(\nu_A-\frac12\right)}\right)\mathds1\mp\frac{i}{2}\Omega
\geq0& \text{if }& \nu_A\geq \nu_B \\
          &&\\
\nu_B\left(\frac{\nu_A+\frac12}{\nu_A-\nu_B}
\right)
\left(\frac{\nu_B\left(\nu_A+\frac12\right)-\nu_B\nu_A-\frac12\nu_A}{\nu_B\left(\nu_A+\frac12\right)}\right)\mathds1\mp\frac{i}{2}\Omega\leq0& \text{if }& \nu_A\leq \nu_B
    \end{array}\right.&\nonumber\\
 &  \iff \left\{\begin{array}{ccc}
   \frac12  \mathds1\mp\frac{i}{2}\Omega
\geq0& \text{if }& \nu_A\geq \nu_B \\
          &&\\
\frac12\mathds1\pm\frac{i}{2}\Omega\geq0& \text{if }& \nu_A\leq \nu_B
    \end{array}\right.&
\end{align}
which is always verified since $\frac12\mathds1$ is the covariance matrix of the vacuum state, which of course satisfies the uncertainty principle.

\section{Definition of standard distances between quantum states}
\label{Appendix:defDistances}

The trace distance \cite{NielsenChuang2000} between two states is defined as
\begin{equation}
D_{\text{trace}}(\rho_A, \rho_B) = \frac{1}{2} \mathrm{Tr} \left| \rho_A - \rho_B \right|=\frac12\tr\sqrt{(\rho_A-\rho_B)^\dag(\rho_A-\rho_B)}.
\end{equation}It is symmetric, invariant under unitary transformations, and is a metric, in the mathematical sense. It gives 0 if and only if the states are identical and 1 if and only if the states are orthogonal. It is also related to the Helstrom bound \cite{Helstrom1} that gives the minimum error probability when distinguishing two quantum states. However, there is no general formula for Gaussian states. Estimates for Gaussian states where recently found in \cite{Bittel2025optimalestimatesof} and it is known that for all states, the trace distance is bounded by the fidelity as
\eq{1-F(\rho_A,\rho_B)\leq D_{\rm{trace}}(\rho_A,\rho_B)\leq\sqrt{1-F(\rho_A,\rho_B)^2}.}
Here, the fidelity \cite{UHLMANN1976273,Jozsa01121994,Banchi,PhysRevA.73.062330,PhysRevA.71.032336} is a measure of similarity between two states and is given by 
 \eq{F(\rho_A, \rho_B) = \left( \mathrm{Tr} \sqrt{\sqrt{\rho_A} \rho_B \sqrt{\rho_A}} \right)^2,}
  with values ranging from 0 to 1. It is symmetric, gives 1 if and only if the states are identical and is invariant under unitary transformations. It does not, however satisfiy the triangle inequality and as such is not a proper metric.
To convert fidelity into a proper metric, we use the Bures distance \cite{Bures1969,PhysRevA.77.062319,Holevo2024} defined as 
  \begin{equation}
D_{\text{Bures}}(\rho_A, \rho_B) = \sqrt{2 - 2\sqrt{F(\rho_A, \rho_B)}}.
\end{equation}
  The overlap $\mathcal{O}(\rho_A,\rho_B)=\tr[\rho_A\rho_B]$ is another measure of similarity that reduces to the fidelity for pure states. For Gaussian states it is easily computed as the integral of two Gaussian Wigner functions.
Another well-used measure in quantum state discrimination is the quantum relative entropy \cite{Petz1986,MuellerLennert2013,Wilde2014,Parthasarathy,Seshadreesan}. Extending the classical Kullback-Leibler divergence to the quantum setting, it is defined as \begin{equation}
S(\rho_A \| \rho_B) = \mathrm{Tr} \left( \rho_A \log \rho_A - \rho_A \log \rho_B \right).
\end{equation} and can be seen as a special case of the Petz–Rényi \cite{Petz1986} and sandwich \cite{MuellerLennert2013,Wilde2014} relative Rényi entropies with $\alpha=1$. In this context, note that the log of the fidelity is then a special case of the sandwich  Rényi relative entropy with $\alpha=1/2$ \cite{Seshadreesan}. The relative Rényi entropy is non-negative and vanishes if and only if $\rho_A=\rho_B$, but note that it is asymmetric in its arguments.
Finally, we consider the Hilbert–Schmidt distance \cite{OZAWA2000158,Coles} defined as 
\begin{equation}
D_{\text{HS}}(\rho_1, \rho_2) = \sqrt{\mathrm{Tr}[ \left( \rho_1 - \rho_2 \right)^2]}.
\end{equation}
Contrary to the trace distance, the Hilbert-Schmidt distance does not have a direct operational meaning \cite{OZAWA2000158}, but it has the advantage of being computable without doing matrix diagonalization. For general Gaussian states, the Hilbert-Schmidt distance can be computed with the overlap of 2 Gaussian states, using the fact that $\tr[(\rho_1-\rho_2)^2]=\tr[\rho_1^2]-2\tr[\rho_1\rho_2]+\tr[\rho_2^2]$.

\section{Inequality between Bures and Wasserstein distances for thermal states.}
\label{Appendix:ineqaulityBuresWasserstein}
Let us prove Eq.~\eqref{eq:inewBuresWasserstein} for two thermal states with covariance matrices $A=\nu_A\mathds{1},\ B=\nu_B\mathds{1}$
such that $\nu_A\geq\nu_B$.
The Wasserstein distance is given by 
\begin{equation}
D^2(\rho_A,\rho_B)=\nu_A+\nu_B- \sqrt{(2\nu_B-1)(2\nu_A+1)}=\nu_A+\nu_B- \sqrt{4\nu_A\nu_B-1-2\nu_A+2\nu_B}.
\end{equation}
On the other hand, we can compute 
\begin{align}
\frac{(\nu_A\!+\!\nu_B)}{2}D^2_{\rm{Bures}}(\rho_A,\rho_B)&=
(\nu_A+\nu_B)\left(1- \sqrt{\frac{\sqrt{\left(4 \nu_A^2-1\right) \left(4 \nu_B^2-1\right)}+(4 \nu_A \nu_B+1)}{\sqrt{2}(\nu_A+\nu_B)^2}}\right)\nonumber\\
% &=\nu_A+\nu_B- \sqrt{\frac{\sqrt{\left(4 \nu_A^2-1\right) \left(4 \nu_B^2-1\right)}+(4 \nu_A \nu_B+1)}{\sqrt{2}}}\\
&=\nu_A\!+\!\nu_B\!- \sqrt{\!\frac{\sqrt{\left(4 \nu_A^2\!-\!1\right) \left(4 \nu_B^2\!-\!1\right)}+2+2\nu_A-2\nu_B+(2\nu_B\!-\!1)(2\nu_A\!+\!1)}{\sqrt{2}}}\nonumber\\
&=\nu_A\!+\!\nu_B\!- \sqrt{(2\nu_B\!-\!1)(2\nu_A\!+\!1)}\sqrt{\frac{\frac{\sqrt{\left(4 \nu_A^2-1\right) \left(4 \nu_B^2-1\right)}}{(2\nu_B-1)(2\nu_A+1)}+\frac{2+2\nu_A-2\nu_B}{(2\nu_B-1)(2\nu_A+1)}+1}{\sqrt{2}}}\nonumber\\
&\leq\nu_A+\nu_B- \sqrt{(2\nu_B-1)(2\nu_A+1)}\nonumber\\
&=D^2(\rho_A,\rho_B)
\end{align}
where we used the fact that
\begin{align}
    \sqrt{\frac{\frac{\sqrt{\left(4 \nu_A^2-1\right) \left(4 \nu_B^2-1\right)}}{(2\nu_B-1)(2\nu_A+1)}+\frac{2+2\nu_A-2\nu_B}{(2\nu_B-1)(2\nu_A+1)}+1}{\sqrt{2}}}&=\frac{1}{2^{1/4}}\sqrt{\underbrace{\sqrt{\frac{\left(2 \nu_A\!-\!1\right) \left(2 \nu_B\!+\!1\right)}{(2\nu_B\!-\!1)(2\nu_A\!+\!1)}}}_{\geq1}+\underbrace{\frac{2+2\nu_A\!-\!2\nu_B}{(2\nu_B\!-\!1)(2\nu_A\!+\!1)}}_{\geq0}+1}\nonumber\\
    &\geq \frac{2^{1/2}}{2^{1/4}}\geq1
\end{align}
because
\begin{align}
    \frac{\left(2 \nu_A-1\right) \left(2 \nu_B+1\right)}{(2\nu_B-1)(2\nu_A+1)}\geq 1 &\iff \ \left(2 \nu_A-1\right) \left(2 \nu_B+1\right)\geq(2\nu_B-1)(2\nu_A+1)\nonumber\\&\iff \nu_A-\nu_B\geq\nu_B-\nu_A\iff\nu_A\geq\nu_B.
\end{align}

\end{document}